\def\year{2022}\relax
\documentclass[letterpaper]{article} 
\usepackage{aaai22}  
\usepackage{times}  
\usepackage{helvet}  
\usepackage{courier}  
\usepackage[hyphens]{url}  
\usepackage{graphicx} 
\usepackage{natbib}  
\usepackage{caption} 
\DeclareCaptionStyle{ruled}{labelfont=normalfont,labelsep=colon,strut=off} 
\usepackage{algorithm}
\usepackage{algorithmic}

\usepackage{newfloat}
\usepackage{listings}
\floatstyle{ruled}
\newfloat{listing}{tb}{lst}{}
\floatname{listing}{Listing}
\usepackage{xcolor}
\usepackage{gensymb}

\title{PDBench: Evaluating Computational Methods for Protein Sequence Design}

\author {
 Leonardo V. Castorina\equalcontrib\textsuperscript{\rm 1},
 Rokas Petrenas\equalcontrib\textsuperscript{\rm 2},
 Kartic Subr\textsuperscript{\rm 1}
 and Christopher W. Wood\thanks{To whom correspondence should be addressed.}\textsuperscript{\rm 2}
}
\affiliations {
 \textsuperscript{\rm 1} School of Informatics, University of Edinburgh, 10 Crichton St, Newington, Edinburgh EH8 9AB\\
 \textsuperscript{\rm 2} School of Biological Sciences, University of Edinburgh, Roger Land Building, Edinburgh EH9 3FF\\
}

\begin{document}

\maketitle

\begin{abstract}
Proteins perform critical processes in all living systems: converting solar energy into chemical energy, replicating DNA, as the basis of highly performant materials, sensing and much more. While an incredible range of functionality has been sampled in nature, it accounts for a tiny fraction of the possible protein universe. If we could tap into this pool of unexplored protein structures, we could search for novel proteins with useful properties that we could apply to tackle the environmental and medical challenges facing humanity. This is the purpose of \emph{de novo} protein design.

Sequence design is an important aspect of \emph{de novo} protein design, and many successful methods to do this have been developed. Recently, deep-learning methods that frame it as a classification problem have emerged as a powerful approach. Beyond their reported improvement in performance, their primary advantage over physics-based methods is that the computational burden is shifted from the user to the developers, thereby increasing accessibility to the design method. Despite this trend, the tools for assessment and comparison of such models remain quite generic. The goal of this paper is to both address the timely problem of evaluation and to shine a spotlight, within the Machine Learning community, on specific assessment criteria that will accelerate impact.

We present a carefully curated benchmark set of proteins and propose a number of standard tests to assess the performance of deep learning based methods. Our robust benchmark provides biological insight into the behaviour of sequence-design methods, which is essential for evaluating their performance and practical utility. We compare five existing models with two novel models for sequence prediction. Finally, we test the designs produced by these models with AlphaFold2, a state-of-the-art structure-prediction algorithm, to determine whether they are likely to fold into the intended 3-Dimensional shapes.

\end{abstract}


\section{Background}
\begin{figure*}
\centering
\includegraphics[width=.9\linewidth]{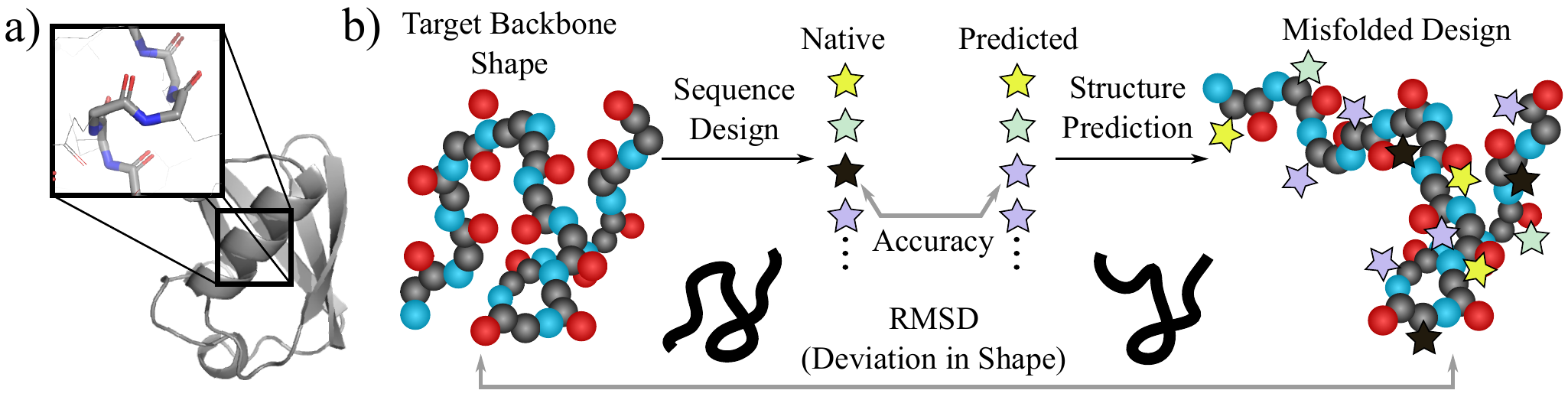}
\caption{\label{fig:overview}
a) An experimentally determined protein structure (PDB ID: 1ubq) rendered in ``cartoon'' format using PyMol \cite{delano_pymol_2002}. The box show atomic level detail with the backbone (tubes) and side chains (lines). Nitrogen atoms are in blue, carbon atoms in grey and oxygen atoms in red. b) A cartoon of protein design, which aims to identify sequences of amino acids (stars) that are able to produce a target backbone shape (circles). This is commonly posed as an inference problem and errors are measured using aggregate metrics such as accuracy of the inferred sequences. Our experiments reveal that measuring the deviation (RMSD) of the designed protein, as predicted by AlphaFold2~\cite{Jumper2021}, from the target structure provides more insight than generic metrics.
}

\end{figure*}

Proteins are the molecules that perform almost all of the biochemical work in all living things. They have a staggering array of functionality from incredibly performant materials, like silks and wools, to some of the most efficient catalysts, capable of accelerating complex chemical reactions \cite{alberts}. Beyond their roles in nature, proteins are broadly applied in industry and medicine. They are the active ingredient in many recent blockbuster drugs, such as immunotherapies for cancer \cite{Nahta2006}, and protein catalysts (enzymes) are increasingly used for chemical synthesis, providing greener alternatives to traditional chemical processes \cite{Wu2020}.

Proteins are formed from long polymers known as polypeptides. Each polypeptide is assembled from fundamental building blocks called amino acids. These building blocks have a common element that is bonded together to make the polypeptide chain called the backbone, and a variable region called the side chain (see Fig.~\ref{fig:overview}), which has 20 possible chemical groups \cite{alberts}. The sequence of amino acids (primary structure) leads to the formation of local and long-range chemical interactions (secondary and tertiary structure respectively), which induce the polypeptide to form a distinct 3D structure, and it is this structure that leads to the proteins function \cite{Anfinsen1961}. Polypeptides vary in length, from small peptides composed of only a few amino acids to huge proteins containing tens of thousands of amino acids, with a median length of around 300 \cite{Brocchieri2005}. The complexity is further increased as multiple polypeptides can assemble to form larger structures (quaternary structure).

The 3D structure of proteins can be determined using 3 main experimental methods: X-ray crystallography, NMR spectroscopy and Cryo-Electron Microscopy. Advances in these techniques have led to a rapid increase in the amount of structural data that is available, with $>180,000$ structures available in the Protein Data Bank (PDB) \cite{Berman2000} as of August 2021. The structures produced by the diligent work of thousands of structural biologists, combined with the foresight and infrastructure to openly shared these data, has given us a deep insight into how proteins function. This data set is now rich enough that we can start addressing the problem of how the sequence of amino acids relates to a folded, functional 3D structure, \emph{i.e.} the “Protein Folding Problem”, through computational means. This is an important problem to address as it is far easier to obtain sequence information than it is to obtain structural information, with genomes available for $> 64,000$ organisms in the NCBI database (https://www.ncbi.nlm.nih.gov/genome/), each of which contains the amino-acid sequence of 100s to 10,000s of proteins.

There has been steady improvement in protein-structure prediction algorithms over the past 2 decades \cite{Kryshtafovych2019}, but recent deep-learning based methods have utterly transformed the field, with state-of-the-art methods producing models that are within the experimental error of the structure-determination method \cite{Baek2021,Jumper2021}. The front runner in these algorithms is DeepMind’s AlphaFold2 (AF2), which has now been applied to predict the structure of every protein in the human genome, as well as the genomes of 20 model organisms \cite{Tunyasuvunakool2021}, which provides an invaluable resource for understanding the function of natural proteins.

However, the protein sequences that have been sampled in nature account for a tiny fraction of all the possible proteins. Even for a relatively small protein with around 200 amino acids, there are around 10$^{260}$ possible sequences, which is significantly more sequences than have been sampled in every cell of every organism since proteins arose \cite{BakerBiophys2019}. This means that it is a statistical certainty that the most performant materials, therapeutics and enzymes for any application have not been sampled. If we are to unlock the potential of this pool of unobserved proteins, known as the dark matter of protein folding space \cite{Taylor2009}, we must solve the “Inverse Protein Folding Problem”, that is if we have a desired 3D structure with a useful function, how can we design an amino-acid sequence that will reliably fold into this structure.

To address this challenge, many successful approaches for designing proteins have been developed, including minimal design and rational design, but computational protein design (CPD) has quickly become the most widely used method \cite{Woolfson2021}. The most successful method in this area is Rosetta~\cite{Guntas2010}, a software suite for computational protein design, which uses a physics-based design method, but deep-learning based methods show a lot of promise for a variety of tasks in CPD, in particular fixed-backbone sequence design, where a desired backbone structure is passed as the input and an amino acid sequence is returned as the output~\cite{zhang_prodconn_2020, Qi2020}.

Current methods of benchmarking fixed-backbone sequence design focus on sequence recovery, where the backbones of natural proteins with known amino-acid sequences are passed as the input and the accuracy of the method is measured by the degree of identity between the predicted sequence and the true sequence \cite{zhang_prodconn_2020,Qi2020,strokach_fast_2020}. However, this approach is fundamentally flawed, and accuracy values of this type do not reflect the real world utility of the design methodology.

To take an extreme example, if a sequence design method had 100\% sequence recovery, this would not necessarily be a useful feature. Very often the purpose of protein design is to diversify the properties of known proteins to make them better suited to a particular application, for example stability or expressability \cite{Goldenzweig2016}. With a sequence recovery of 100\%, no diversity would be generated, and the method may not be capturing the functional redundancy that we observe in natural proteins, that is, many sequences can fold to adopt the same overall shape (fold). Ultimately, \textit{we must move beyond simplistic methods for evaluating design methodologies} and provide information to users that will help them to assess whether a specific method will be appropriate for their target application.

\begin{figure*}[htbp]
\includegraphics[width=\linewidth]{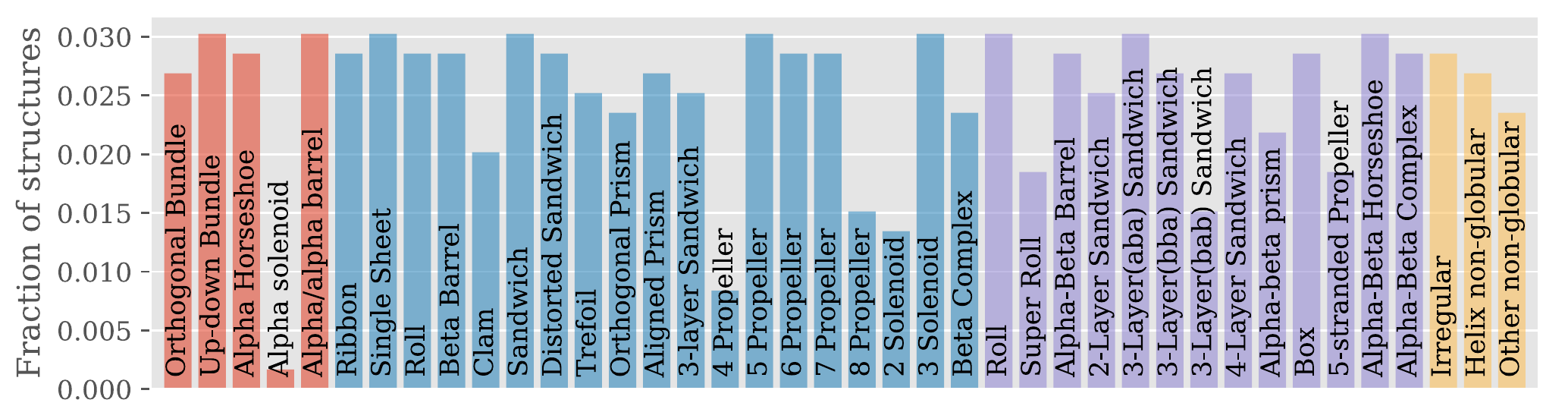}
\caption{\label{fig:benchset}
Benchmark set composition. 40 protein architectures grouped into 4 categories: mainly-$\alpha$ (red) - 70 chains, mainly-$\beta$ (blue) - 282 chains, $\alpha$-$\beta$ (purple) - 196 chains, special (yellow) - 47 chains. 
}
\end{figure*}

\subsubsection{Our Contributions} 

Here, we have created a benchmark for evaluating the performance of CPD methods, which gives a more holistic view of performance. It combines a set of about 600 structures carefully sampled from a broad range of protein folds, with diverse secondary structures compositions and resolutions. We applied this benchmark to evaluate state-of-the-art methods for sequence design, both physics- and neural-network based, and uncovered useful information about their applicability and performance. Finally, we used this benchmark to guide the development of highly performant convolutional, graph and hybrid CNN/GNN methods, and tested their performance by applying AF2 to predict the structure of our designed sequences. Our results show that about 85\% of our designs are predicted to fold into their intended 3D structure (within 3 \r{A}), despite having $<$ 42\% sequence identity with the native sequence, which demonstrates that they have clear utility to design novel proteins with diverse properties.

\section{PDBench: Evaluation benchmark set}\label{eval_benchmark}
\subsubsection{Diverse Set of Structures} 
\begin{figure}[htbp]
\centering
    \begin{tabular}{@{}c@{}c@{}}
        \includegraphics[width=1\linewidth]{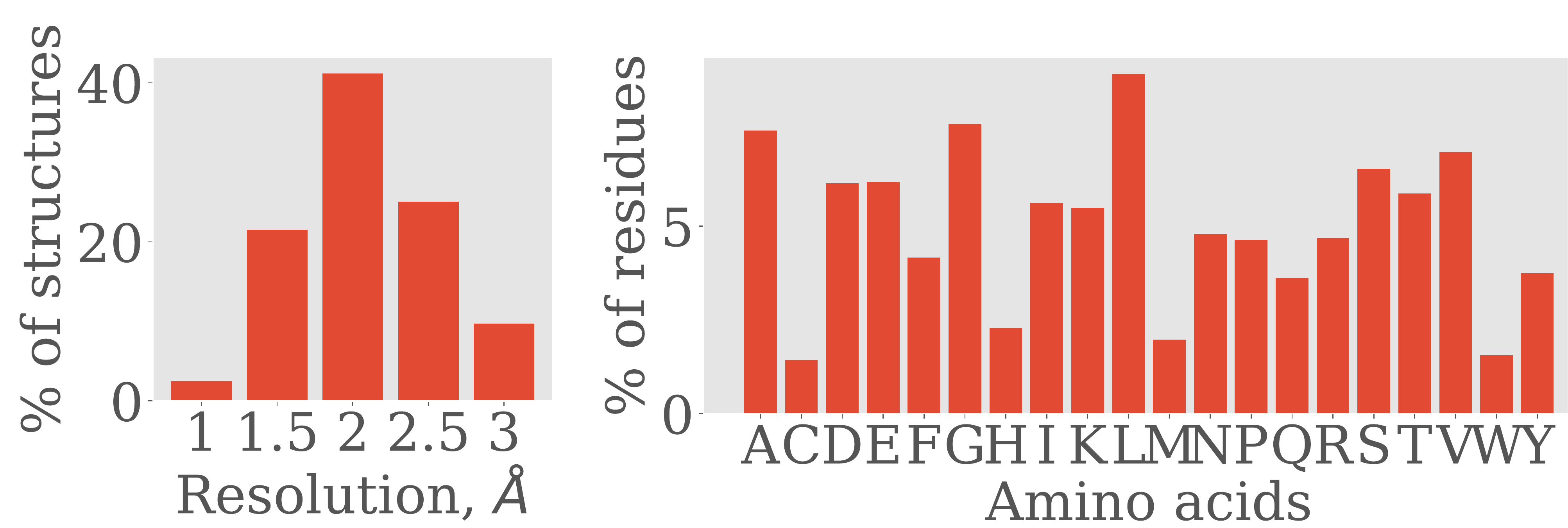}
    \end{tabular}
    \caption{\label{fig:bench_dist} Resolution and amino acid distribution in the benchmark set.
    }
\end{figure}

Our benchmark set contains 595 protein structures spanning 40 protein architectures that are clustered into 4 fold classes (see Fig.~\ref{fig:benchset}): mainly-$\alpha$, mainly-$\beta$, $\alpha$-$\beta$ and special, as presented in the CATH database \cite{knudsen_cath_2010}. The `special' category contains proteins that do not have regular secondary structure.
Crystal structures with maximum resolution of 3 Å and up to 90\% sequence identity were carefully chosen to cover the structural diversity present in the PDB (see Fig.~\ref{fig:benchset}). This ensures that the performance is evaluated on high- and low-quality inputs (see Fig.~\ref{fig:bench_dist}) and the results are not biased towards the most common protein architectures. The benchmark structures were prepared by removing all non-backbone atoms using AMPAL \cite{wood_isambard_2017}.

\subsubsection{Benchmarking Tool} 
We have developed an open-source benchmarking library that is implemented in Python. The inputs to our program are a prediction matrix (in .csv format) and a dataset map (in .txt format). 
The prediction matrix is $n\times20$ for a protein with $n$ amino acids in the polypeptide chain. Each row in the matrix encodes prediction probabilities across the 20 canonical amino-acid classes. The dataset map contains a list of protein chains to be evaluated with the benchmark. The outputs of our program are the metrics for each model in a plot, as well as the option to generate comparison plots between different models to compare their performances. The software uses the AMPAL library to read protein sequences and then replaces non-canonical residues with standard amino acids residues for compatibility with protein design software \cite{wood_isambard_2017}. For example, selenomethionine is converted to methionine. Optionally, non-canonical residues can be omitted from calculations. 

DSSP is used to assign the secondary structure for each residue \cite{Joosten2010}. For simplicity, predictions for $\alpha$-helix, $\pi$-helix and 3$_{10}$-helix are combined and treated as helices. Predictions for hydrogen-bonded turn, bend and isolated $\beta$-bridge residues are combined and treated as structured loops. CATH database \cite{knudsen_cath_2010} is used to assign protein architectures to chains.

\subsubsection{Metrics}
We calculate four groups of metrics: 1) recall, precision, AUC, F1 score, Shannon's entropy, and prediction bias \emph{for each amino acid class}; 2) accuracy, macro-precision, macro-recall, similarity and top-3 accuracy \emph{for each protein chain}; 3) accuracy, macro-precision, macro-recall, similarity and top-3 accuracy \emph{for each secondary structure type};  4)  accuracy, macro-precision, macro-recall, similarity and top-3 accuracy \emph{for each protein architecture}. All metrics except similarity and prediction bias are calculated with SciKit-Learn \cite{pedregosa_scikit-learn_2011}. Prediction bias is a metric measuring the discrepancy between the occurrence of a residue and the number of times it is predicted.

Accuracy-based metrics are useful, but there is functional redundancy between amino acids, as many side chains have similar chemistry. The similarity of amino acids can be determined by the relative frequency of substitution of one amino acid for another observed in natural structures, using substitution matrices such as BLOSUM62, which we combine into a similarity score for the sequence \cite{Henikoff1992}. We also created torsion angle comparison plots between true and predicted residues for each model. $\Psi$ and $\phi$ angles for the plots are extracted using AMPAL \cite{wood_isambard_2017}; the residue count for each torsion angle pair for each amino acid is normalized by the true number of that amino acid in the dataset. 

\subsubsection{Structure Evaluation}
We computationally validate designed sequences, using AF2 \cite{Jumper2021}, to determine if the sequence adopts the intended 3D structure. The sequence predictions are used as an input and the predicted structure is compared to the target structure, calculating a length-normalised form of Root Mean Squared Deviation (RMSD) \cite{Carugo2008}. As the folding predictions are computationally demanding, we randomly chose 59 monomeric\footnote{ AF2 performance is higher in protein monomers compared to multimeric complexes \cite{Jumper2021}} protein structures from our benchmark. We used a modified version of ColabFold~\cite{sergey_ovchinnikov_2021_5123297} to allow for multiple simultaneous predictions on a single GPU (about 2 days for each set of 59 structures). RMSD was calculated using the align command in PyMOL  \cite{delano_pymol_2002}.

\section{Models evaluated}\label{existing_models}


\subsection{Physics-based Models} 
We tested two state-of-the-art physics-based methods: \textbf{EvoEF2}~\cite{Huang2019} and \textbf{Rosetta}~\cite{alford_rosetta_2017}. Fixed backbone sequence design protocols were performed, details of which can be found in the Supplementary Material.




\subsection{Existing Deep Learning-based Models}
The input to CNN models is a fixed-size 3D grid, that we call a \emph{frame}, which is centred on the C$\alpha$ atom of the input amino acid, with a frame for each residue in the protein sequence. The network is trained to classify, predicting probabilities across a fixed set of $20$ classes for each frame. The input to GNN models on the other hand is the whole protein as a graph. Each residue is represented as a node in the graph, with an edge connecting two nodes if the corresponding residues are within a predefined distance threshold. The GNN is then trained to produce a 20-dimensional embedding that represents the class probabilities for all residues at once. An illustration of the different types of data input is shown in the Supplementary Materials.

\subsubsection{ProDCoNN (CNN)} We replicated the architecture described in \citet{zhang_prodconn_2020}. We used our open-source voxelisation library to create the protein dataset used to train the model. The neural network was built using Keras \cite{chollet2015keras}.

\subsubsection{DenseCPD (CNN)} \citet{Qi2020} proposed using a 3D DenseNet architecture \cite{huang2019convolutional} for sequence design. Their model contains 3M trainable parameters, however we were not able to replicate it with the information given. \citet{Qi2020} kindly shared their model architecture for us to train with the same dataset as the other models, although the architecture provided was different to that described in the paper. When trained with the same data as the other models, we were not able to match the published performance of DenseCPD, therefore we used their trained model. \emph{However, it must be noted that this model has been trained on some of the benchmark structures}.

\subsubsection{ProteinSolver (GNN)} is a model proposed by \citet{strokach_fast_2020}. Their code and the model are open-source and available online. The model uses a threshold of 12 \r{A} to consider two residues connected by an edge.

\subsection{Novel models} \label{our_models}
\subsubsection{TIMED-Unbalanced (CNN)} 
We used the Keras framework as a high-level interface to TensorFlow \cite{tensorflow_developers_tensorflow_2021}, to create a deep CNN, which we refer to as TIMED (Three-dimensional Inference Method for Efficient Design). Details of the architecture are in the Supplementary Material. One of the key features of our neural network is the use of the final Global Average Pooling (GAP) layer rather than a Fully Connected (Dense) layer, to preserve spatial information \cite{lin_network_2014}. The model also uses Spatial Dropout rather than standard dropout to help enforce this relationship. The model was trained with categorical cross entropy loss.

\subsubsection{TIMED-Balanced (CNN)} 
 To avoid biasing our neural network towards the natural frequency of amino acid, we use a random undersampling method. In the training and validation sets, all residue classes are capped to match the number of the least abundant residue. At the beginning of every epoch, for residues with a higher count than the minimum, frames are randomly re-sampled to increase the number of total residues observed by the network.

\subsubsection{GX (GNN + GNN-CNN Hybrid)} Our Graph eXpanded method involves a GNN architecture with several SAGE layers \cite{hamilton2018inductive} and a mean aggregation function. While we experimented with several features, we utilises the frame predictions of TIMED as its input features and the coordinates of atoms C, N, O and C$\alpha$. We call this GX[PC] since it is a hybrid CNN-GNN model which aim at improving on the high CNN performance by adding longer-range information captured by the GNN. We used a weighted categorical cross entropy loss where weights were obtained from the inverse natural frequency of residues \cite{bairoch_2013}.

\section{Experimental results}

\subsection{Implementation details}
The TIMED model (CNN) was built with Keras. We trained our model for 50 epochs which took about 49 hours.
The GX model (GNN) was built with PyTorch \cite{NEURIPS2019_9015}. Each batch of the dataset contained 200 structures for the CSD3 server and 1,000-1,200 structures for our internal server. Training took about 10 hours for about 1000 epochs.
The hardware used for training was a combination of the Cambridge Service for Data-Driven Discovery (CSD3) (NVIDIA Tesla P100 16GB GPU and 36 cores) and our internal servers (Intel Core i9-10980XE CPU @ 3.00GHz (36 cores) and NVIDIA Quadro RTX 8000 48GB). Weights \& Biases was used to track experiments and for Bayesian hyper-parameter sweep \cite{wandb}.

\begin{figure}[H]
\centering
    \begin{tabular}{@{}c@{}}
        \includegraphics[width=\linewidth]{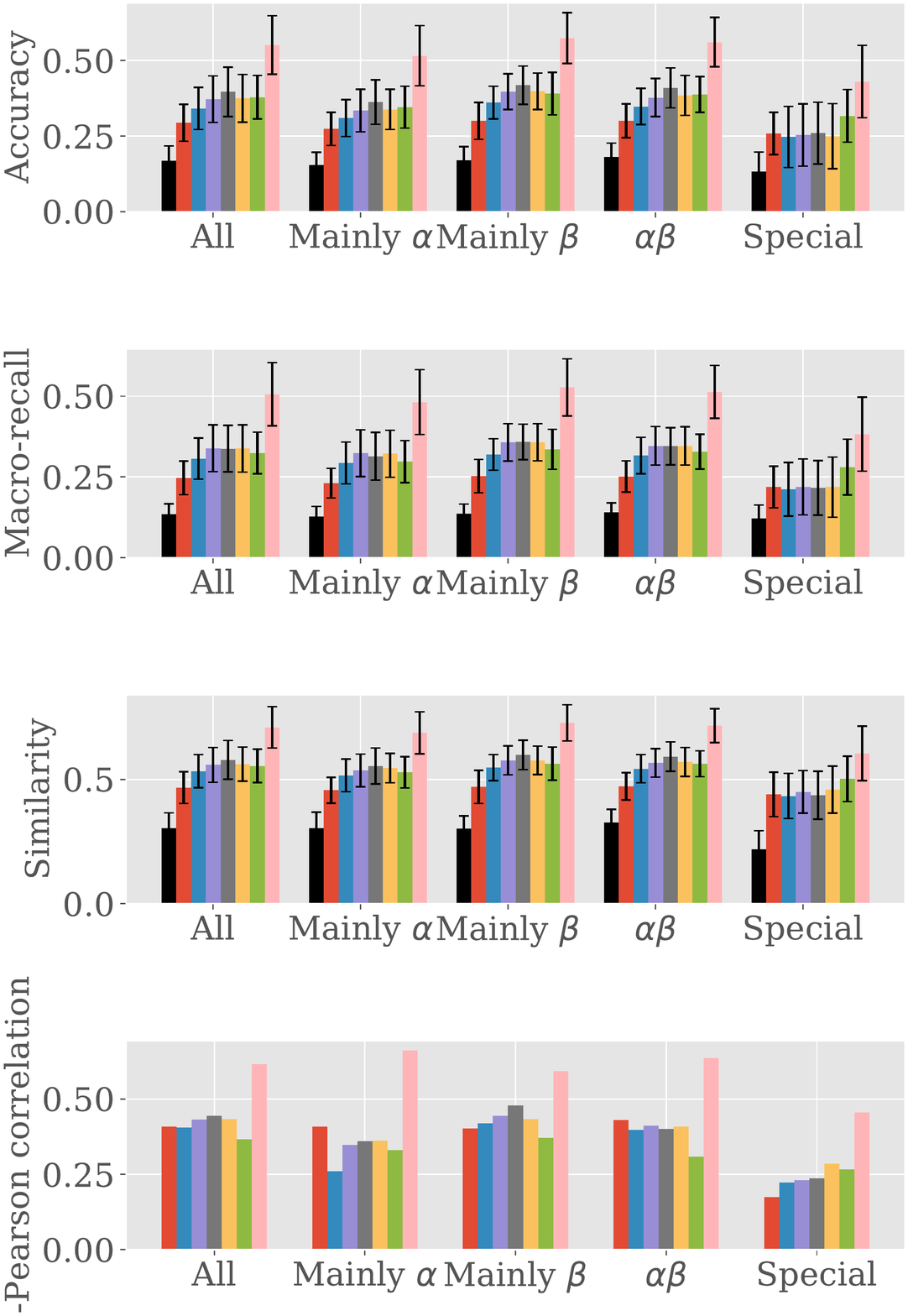} \\ 
        \includegraphics[width=\linewidth]{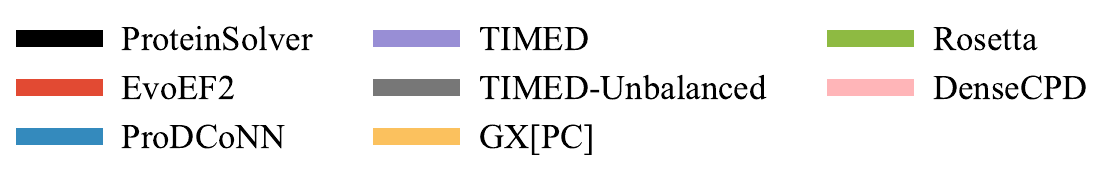}
    \end{tabular}
    \caption{\label{fig:combined_perf_corr}
   The first three plots compare performances  of models across classes of folds. The plot at the bottom shows the negated correlation coefficient  (Y axis) between macro-recall and the resolution (in \r{A}) of the input structure. The performance of DenseCPD depends on input with fine resolution. All p-values were significant ($< 10^{-8}$) except for ProteinSolver (0.3) which is excluded from the plot.
    }
\end{figure}

\subsection{Model Performance per Class}

We compared all the models with our benchmark using the accuracy, macro-recall and similarity metrics (see Fig.~\ref{fig:combined_perf_corr}). Since proteins can adopt a wide range of structures, per-fold metrics provide more meaningful insights about the performance of a model. We divided our benchmark set (595 structures) into four categories of protein folds as explained in Sec. \ref{eval_benchmark}. See Fig.~\ref{fig:benchset} for the proportion of structures in each category. We calculated metrics for each protein structure, and averaged the metrics across folds. Error bars show one standard deviation from the mean within each group. 

Between the physics-based methods, Rosetta outperformed EvoEF2 in all our tests and all metrics by about 4-10 \%. All TIMED and DenseCPD models marginally outperformed the physics-based methods across the metrics regardless of structure types except for the ``special'' class. DenseCPD\footnote{Caveat: Our retrained version of the DenseCPD model was unable to match its claimed performance. We therefore used the DenseCPD model trained and shared kindly by its authors. However, this contained some benchmark structures in its training set.} had the highest performance across all metrics by a considerable margin. ProteinSolver had the lowest performance across all methods (\emph{vide infra}).

  


\subsection{Dependence on Resolution}

The spatial resolutions of experimentally determined protein structures varies by methodology and other factors. The finer the resolution, the more precisely the position of the atoms are known in the structure. We calculated  correlations between macro recall and input resolution (in \r{A}) for all models across different folds (Figure \ref{fig:combined_perf_corr}). Except for ProteinSolver, all models had a significant correlation between resolution and macro-recall, with p-values values of $< 10^{-8}$. DenseCPD consistently had a higher correlation than other models, which could be explained by the fact that it was trained on 2\r{A} structures while the other deep learning models (with the exception of ProteinSolver) were re-trained on 3\r{A} structures. Rosetta exhibited lower correlation than other models on all protein classes except `special'.

\subsection{Prediction Bias}

The composition of the amino acids in proteins is not uniformly distributed, and can vary significantly between different protein folds. Therefore, we investigated the effect of balancing our frame dataset prior to training.  When TIMED and ProDCoNN were trained without balancing, the accuracy on the benchmarking set increased from 38.5\% to 41.1\% and from 35.1\% to 37.7\%, respectively. However, this also resulted in increased prediction bias for the most common amino acids, as well as decreased macro-recall. This is particularly obvious for alanine, glutamate and leucine in $\alpha$-helices (4\%, 10\% and 6\% bias in TIMED; 2\%, 15\% and 10\% in ProDCoNN, respectively) and, to a lesser extent, leucine and valine in $\beta$-sheets (3\% and 5\% in TIMED, 4\% and 7\% in ProDCoNN) 
(see Supplementary Material). When the amino acid distribution was balanced by random undersampling, the prediction bias was closer to 0\% for all amino acids, indicating that predicted sequences had the same amino acid distribution as true sequences. Overall, TIMED architecture appears to be more robust to amino acid imbalance which could be key to designing natural-looking sequences. 

The benchmark also produces torsion angle comparison plots that investigate the of prediction bias in more detail. The geometry of the protein backbone can be described primarily by rotation around the N-C$_{\alpha}$ and C$_{\alpha}$-C bonds. These rotations can be described by the two torsion angles $\Phi$ and $\Psi$. Certain backbone conformations are associated with specific combinations of $\Phi$ and $\Psi$ \emph{e.g.} repeated regions of $\Phi$=-60, $\Psi$=-60 lead to the formation of $\alpha$-helical secondary structure. By comparing torsion-angle frequencies of true and predicted amino acids, we can identify if predicted amino acids are in energetically favourable regions and how their distribution has changed with respect to different structures within proteins. For example, the first iteration of TIMED model had the accuracy of 26\% but 19\% prediction bias for glycine. We discovered that glycine is overpredicted almost everywhere but especially in the  $\alpha$-helical region ($\Psi$=-60, $\Phi$=-60) and, to a lesser extent, in region associated with $\beta$-sheets (around $\Phi$=-140, $\Psi$=140), and is frequently confused for other amino acids  (Fig. \ref{fig:glycine_trap}).

\subsection{Design Evaluation with AlphaFold2 (AF2)}

We randomly selected 59 structures from the benchmark (about 10\% of the benchmark) covering classes of mainly alpha, mainly beta and alpha-beta. We predicted the residue sequences using each of the models in Sec.~\ref{existing_models}. We provided the predictions as input into AF2, a state-of-the-art protein folding model, to obtain hypothetical 3D structures of the predicted sequences. We then compared the RMSD between the predicted structure and the target structure. We did this for each model and for each class. We excluded structures with the ``special'' class as they are highly irregular.

\begin{figure*}[t]
\centering
    \begin{tabular}{@{}c@{}}
        \includegraphics[width=\linewidth]{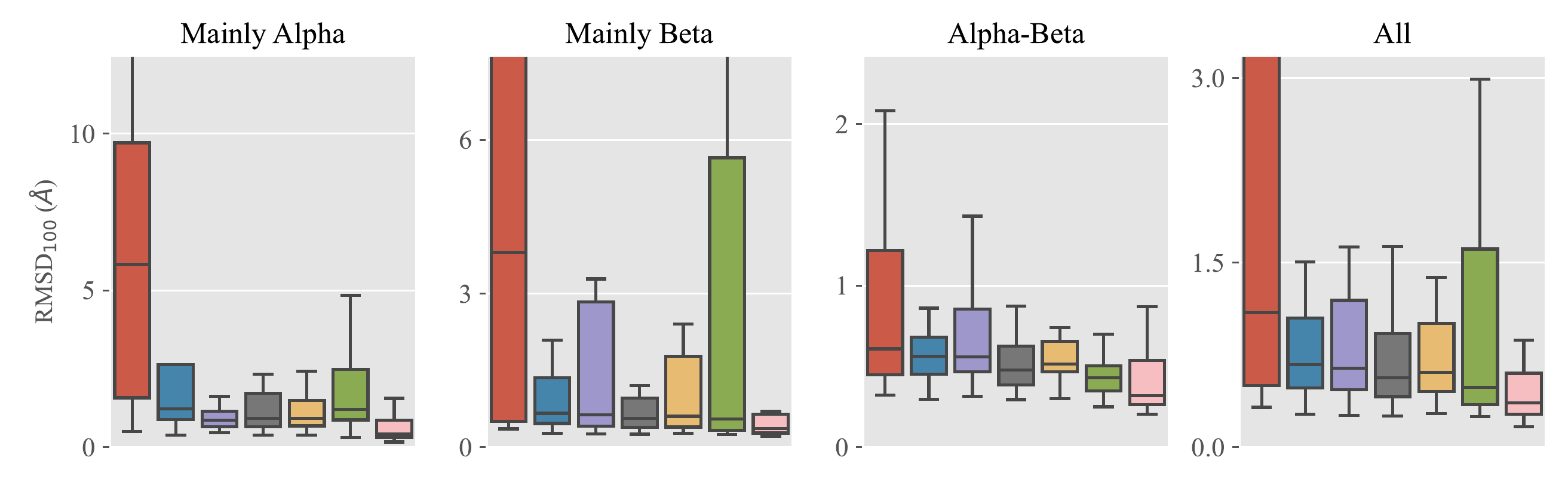}\\
    \includegraphics[width=1\linewidth]{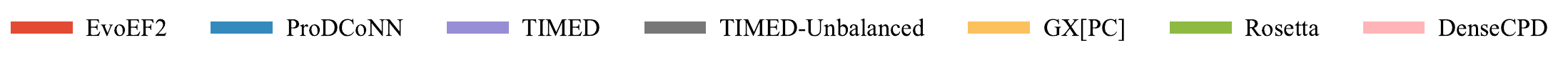}
    \end{tabular}
    \caption{Boxplot showing the ranges of RSMD between the real structure and predicted structure for each model, seperated by protein class. 59 Structures were randomly selected from the benchamrk and the predicted sequence for each model was fed to AF2 to obtain a theoretical structure. The structures were then compared using RMSD$_{100}$, a normalised version of RMSD to compare protein structures of different lengths proposed by \citet{Carugo2008}.}
  \label{fig:af2_rmsd_boxplot}
\end{figure*}

As shown in Fig. \ref{fig:af2_rmsd_boxplot}, the physics-based methods tend to have a larger range of RMSD, especially EvoEF2. Rosetta has a larger RMSD range in the mainly beta fold. Most models performed better for alpha-beta folds. The performance of deep-learning models is usually equivalent or better than the physics-based methods, with DenseCPD performing best overall. It generally has the smallest range in RMSD except for the mainly-alpha fold where it has a larger range of RMSD than TIMED-unbalanced. The unbalanced models seem to outperform the balanced ones in terms of RMSD, consistently over all folds. The GX[PC] model seems to improve the prediction of TIMED-Balanced in the mainly beta fold and alpha-beta. The ProDCoNN model typically tends to perform better in mainly beta structures. 

We also visualise these data using a percentage plot to show how many structures are present at each RMSD threshold (see Supplementary Material). We see that, under 6 \r{A}, deep-learning models tend to have a similar percentage of structures at each RMSD threshold. DenseCPD is the best overall model, but it closely matched by other deep learning models. It is worth reiterating that some of these structure from the benchmark were present training set of DenseCPD. We avoided running ProteinSolver on AF2 given the poor performance in our benchmark. 


%
%
%

\section{Discussion}
\subsubsection{Combinations of Metrics Across Classes}
When considering accuracy metrics (Fig. ~\ref{fig:combined_perf_corr}) along with similarity to the target structures (Fig. ~\ref{fig:af2_rmsd_boxplot}), there is a marked difference in the performance of all the design algorithms across the different fold classes. It is interesting that all of the deep learning based methods performed well when designing ``mainly $\beta$'' structures, as these are traditionally very challenging targets \cite{woolfson_novo_2015,huang_coming_2016}. Furthermore, the accuracy of sequence recovery was more strongly correlated with resolution in the $\beta$-rich classes (``mainly $\beta$ and $\alpha\beta$'') (Fig. ~\ref{fig:combined_perf_corr}), which suggests that the sequence preferences in $\beta$ structure are closely linked to subtle details in the backbone conformation. This might indicate \emph{why} $\beta$ design is more challenging: it is more difficult to generate high-quality backbone models for $\beta$-rich structures compared with $\alpha$-rich structures, most likely due to the higher degree of conformational flexibility compared to $\alpha$ helices, and so our focus should be on improving backbone-modelling techniques for $\beta$ structure rather than sequence-design methods.

\begin{figure}[H]
\centering
    \begin{tabular}{@{}c@{}c@{}}
        \includegraphics[width=.50\linewidth]{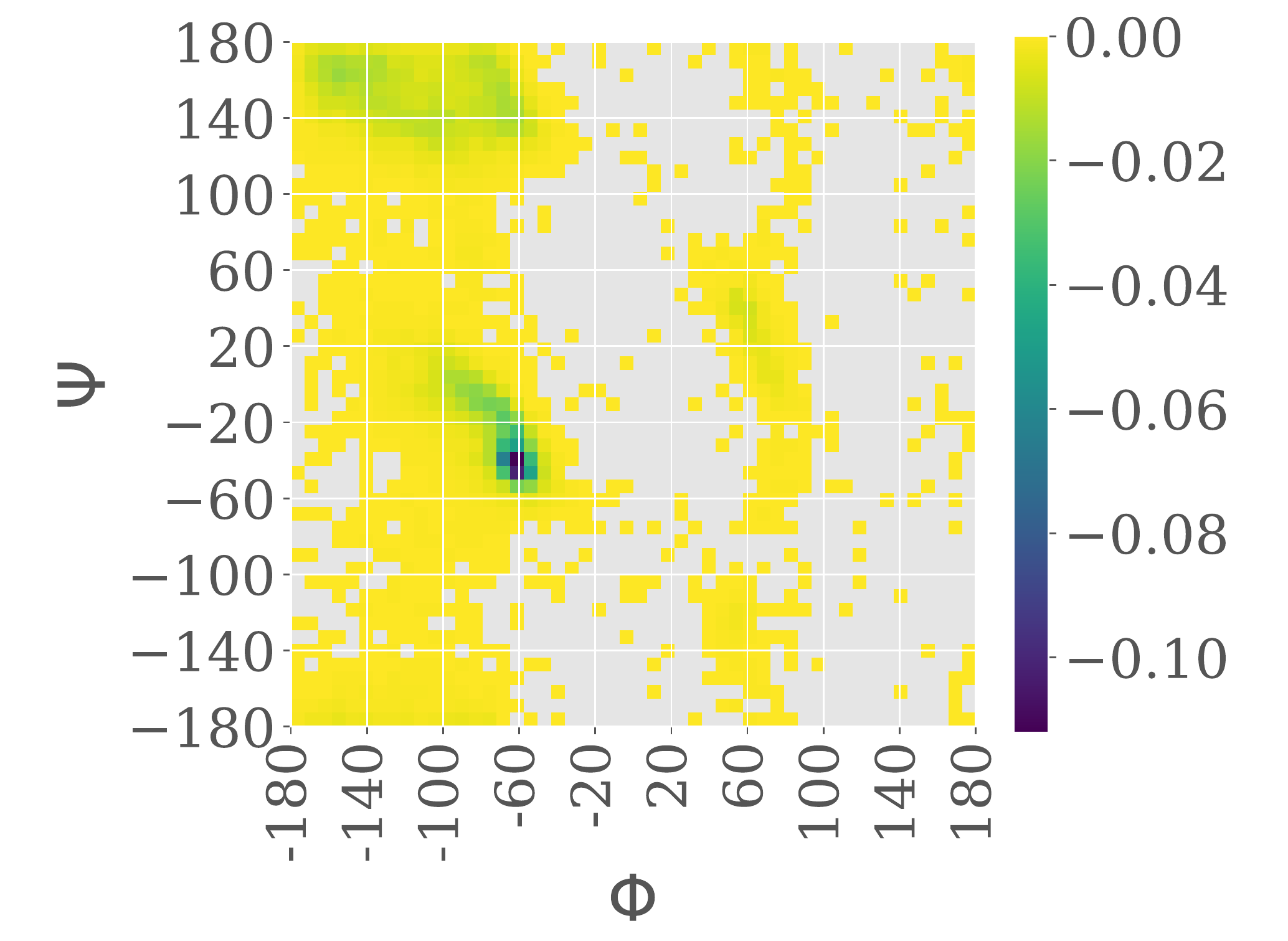}
        \includegraphics[width=.5\linewidth]{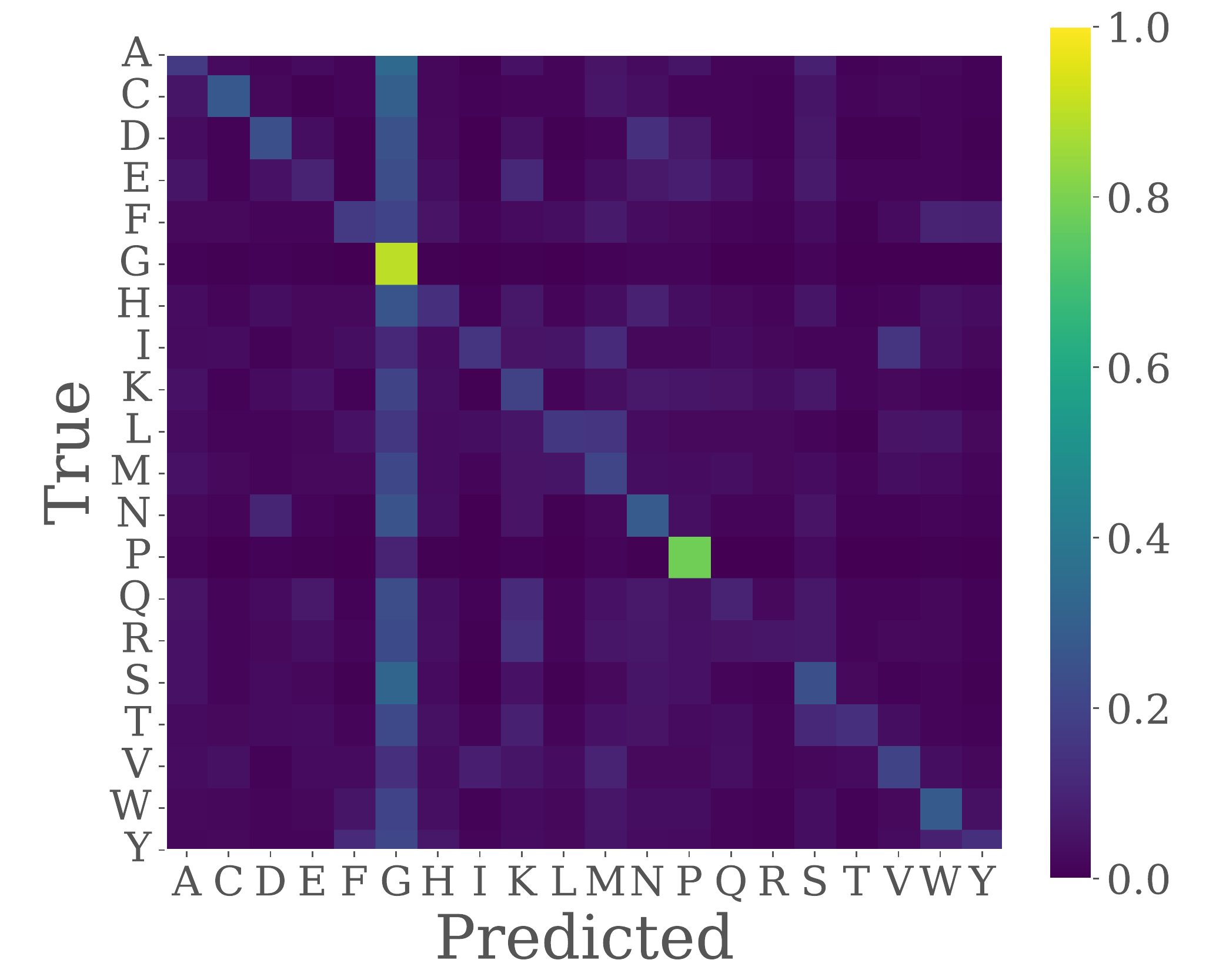}
    \end{tabular}
    \caption{Glycine overprediction. Left: A torsion angle plot showing normalized frequency difference between true and predicted number of glycine amino acids. Negative values indicate increased glycine frequency in predicted sequences. Right: A confusion matrix showing the confusion frequency for amino acid pairs.}
  \label{fig:glycine_trap}
\end{figure}

\subsubsection{Glycine Trap} The over prediction of glycine by our initial model is likely related to the flexibility of this residue, which allows it to adopt a broad combinations of $\Phi$ and $\Psi$ \cite{Lovell2003}. It is well documented that the identity of the side chain has a large impact on the backbone conformation \cite{Shapovalov2011}, therefore it is likely that it is important to the quality of predictions. However, if too much attention is placed on backbone torsion angles, it is easy to imagine that a network would over predict residues with a strong influence on backbone conformation, such as glycine and proline, which is what we see in our initial model (See Fig. \ref{fig:glycine_trap} right). However, glycine is a highly destabilizing $\alpha$-helices \cite{LpezLlano2006}, so it is important to be aware of these biases. PDBench can easily identify cases where this, or other sequence biases, have occurred, which can be used to guide the development of the design methodology.


\subsubsection{Balancing Amino Acids} Using the benchmark, we explored the effect of balancing amino-acid classes. In some ways, unbalanced classes better reflect the biochemical availability of the individual amino acids, which could improve production of the proteins in living systems. However, this would mean that the biases of natural proteins would be reflected in the sequences produced by the design algorithm, when there's strong evidence that functional proteins exist in sequence spaces that are unexplored in nature \cite{Weidmann2019}. The benchmark highlights these sequence biases clearly, and armed with the results of the benchmark, potential users can decide whether this behaviour is desirable for their specific application.

\subsubsection{ProteinSolver and \emph{de novo} Design} We were surprised to see that the performance of ProteinSolver was lower using our benchmark compared to their published results. We hypothesised that this could be due to differences in the input structures. If side-chain atoms of the input structures were included when calculating the distance matrix, which is used as an input to their model, the distance matrix would indirectly encode sequence information present on the backbone scaffold. For example, for long and flexible residues like glutamate and lysine, those residues could be closer to many more residues than the backbone alone, which will have an obvious impact on the distance matrix. When designing \textit{de novo}, the identity of the side-chains is unknown, and so we strip the side chain atoms out of the benchmark structures. We tested ProteinSolver using our benchmark structures both with and without side-chain atoms, which confirmed our hypothesis of label leakage, as the macro-recall dropped from 36.3\% to 13.3\%. ProteinSolver clearly has some utility for making minor changes to natural sequences, but it appears to be unsuitable for \emph{de novo} design.

\subsubsection{Significance Threshold for Metrics} We have discovered that accuracy, macro-recall and other statistical metrics can accurately estimate a model's performance only up to a certain point. For example, the TIMED unbalanced and DenseCPD models differ significantly by accuracy, macro-recall and similarity. However, difference in RMSD in the AlphaFold2 predictions are not significant, so it appears that they both produce sequences that will fold to the target structure. Perhaps, after a certain accuracy point, statistical metrics become less relevant. In the case of \textit{de novo} design for example, high accuracy might limit the utility of the design method, as the sequences produced will have lower variability. In real-world application of protein, the increased diversity of lower accuracy models might be more desirable, especially when the experimental strategy involves high-throughput screening.

\subsubsection{Limitations}
We have used structure predictions from AF2 to validate the designs produced by the sequence design algorithms, which is computationally demanding. To provide context, the evaluation using AF2 (on about 10\% of the benchmark set) took two days of computation on our servers. While the reported performance is impressive \cite{Jumper2021}, it remains to be seen whether this is a useful method for determining if designs will fold in experimental settings. As AF2 is trained on observed protein structures, the predictive power of this model may be lower for sequences outside of observed structural space. In this case, physics-based simulations such as molecular dynamics, might be required to evaluate the AF2 structures further, although these simulations are even more computationally expensive, so reduce the scale at which design can be performed.

\section{Conclusion}
We have presented PDBench, a dataset and software package for evaluating fixed-backbone sequence design algorithms. The structures included in PDBench have been chosen to account for the diversity and quality of observed protein structures, giving a more holistic view of performance. We find that generic metrics such as classification accuracy and recall are less informative when determining the utility of computational models for protein-sequence design, as they can obscure properties of the design method that could have a major impact on the quality of the designed sequences. Finally, structure predictions of designed sequences showed that there appears to be diminishing returns from further improving fixed-backbone design algorithms, and so, in the spirit of quantum physicists before us, perhaps it is time to ``shut up and calculate'' \cite{DavidMermin1989}. We are currently in the process of experimentally validating designs from the TIMED models in the laboratory.

\section{Acknowledgements}
CWW is supported by an Engineering and Physical Sciences Research Council Fellowship (EP/S003002/1). LVC is supported by the United Kingdom Research and Innovation (EP/S02431X/1), UKRI Centre for Doctoral Training in Biomedical AI at the University of Edinburgh, School of Informatics. This work was supported by the Wellcome Trust-University of Edinburgh Institutional Strategic Support Fund (ISSF3). Kartic Subr is supported by a Royal Society University Research Fellowship. 

\bibliography{references}

\end{document}


\maketitle

\section{Guide}

\begin{table}[H]
 \centering

\begin{tabular}{ll}
\textbf{From   Section (Main Paper)} & \textbf{Link to Section / Figure} \\ 
3.1 Physics-based models & Section \ref{physics} \\ 
3.2 Existing Deep Learning-based Models & Figure \ref{fig:data_pipeline} \\ 
3.3 Novel models & Figure \ref{fig:cnn_architecture_fig} \\ 
4.4 Prediction Bias & Figure \ref{fig:prediction_bias_alpha} \\ 
4.5 Design Evaluation with AlphaFold2 & Figure \ref{fig:af2_rmsd_results} and \ref{fig:af2_rmsd_results2} \\ 

\end{tabular}
\end{table}

\section{DenseCPD and TIMED Benchmark Performance with and without train structure}

Our retrained version of DenseCPD was unable to match its claimed performance. We therefore used DenseCPD model trained and shared kindly by its authors. However, this contained some benchmark structures in its training set. We also evaluated denseCPD with training structures excluded from the benchmark set.

\begin{table}[H]
\begin{tabular}{lllllll}
                 & \multicolumn{2}{l}{\textbf{Full benchmark set}} &  &  & \multicolumn{2}{l}{\textbf{denseCPD training   structures excluded}} \\
Metrics          & denseCPD            & TIMED            &  &  & denseCPD                       & TIMED                      \\
Accuracy,   \%   & 56.8                & 38.9             &  &  & 51.6                           & 36.9                       \\
Macro-recall, \% & 53.2                & 37.5             &  &  & 47.8                           & 35.5                       \\
Similarity,   \% & 72.4                & 57.4             &  &  & 68.1                           & 55.5                      
\end{tabular}
\caption{\label{tab:}
 denseCPD performance on the benchmark set with and without training structures. Structures not included in denseCPD training set had on average lower resolution; this resulted in lower metrics not only for denseCPD but also for TIMED. Therefore, denseCPD does not seem to overfit its training data.   }
\end{table}

\section{Models Summary}

We propose two novel models as well as comparing them with other available deep learning- and physics- based methods:
\begin{table}[H]
 \centering
 \begin{tabular}{@{}rccc@{}}
     \multicolumn{1}{c}{\textbf{Method}} & \textbf{Type} & \textbf{Architecture} & \textbf{Novel} \\
     EvoF2 & physics & - & no \\
     Rosetta & physics & - & no \\
     ProdConn & learned & CNN & no \\
     DenseCPD  & learned & CNN & no \\
     ProteinSolver & learned & GNN & no \\
     TIMED & learned & CNN & yes \\
     GX[PC] & learned & GNN & yes 
 \end{tabular}
\caption{\label{tab:ModelList}
 A summary of models compared in this paper.}
 \end{table}

\section{Comparison of two extreme examples from DenseCPD and TIMED}

We then selected two exemplar structures from models DenseCPD and TIMED-Balanced to further understand the performance in the context of protein structures. We selected structures where one model had low RMSD and the other a high RMSD. We selected structures 3CXB and 4EFP.

\begin{figure}[H]
    \centering

    \begin{tabular}{@{}c@{}c@{}}    \includegraphics[width=0.45\linewidth]{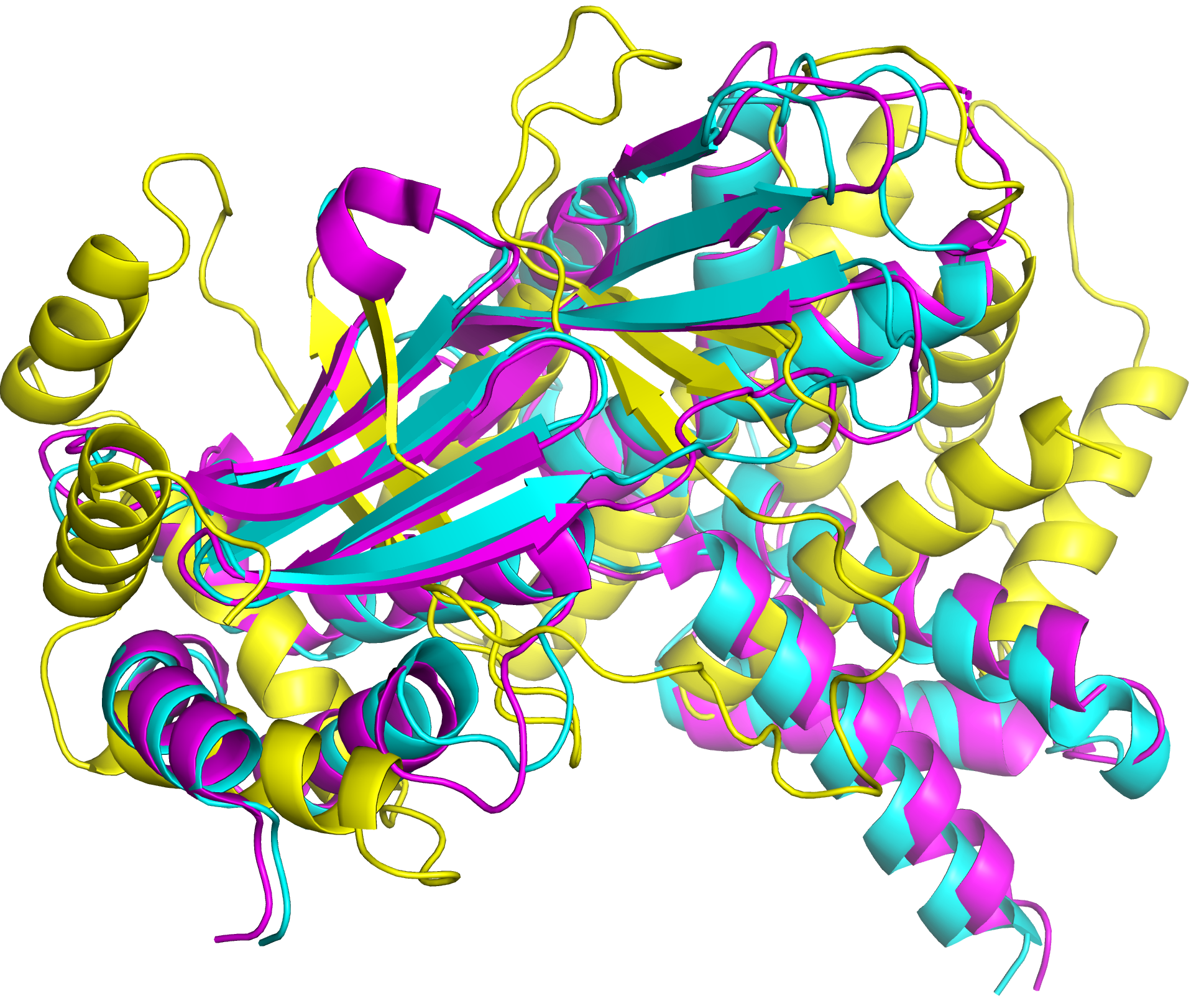} &
    \includegraphics[width=0.45\linewidth]{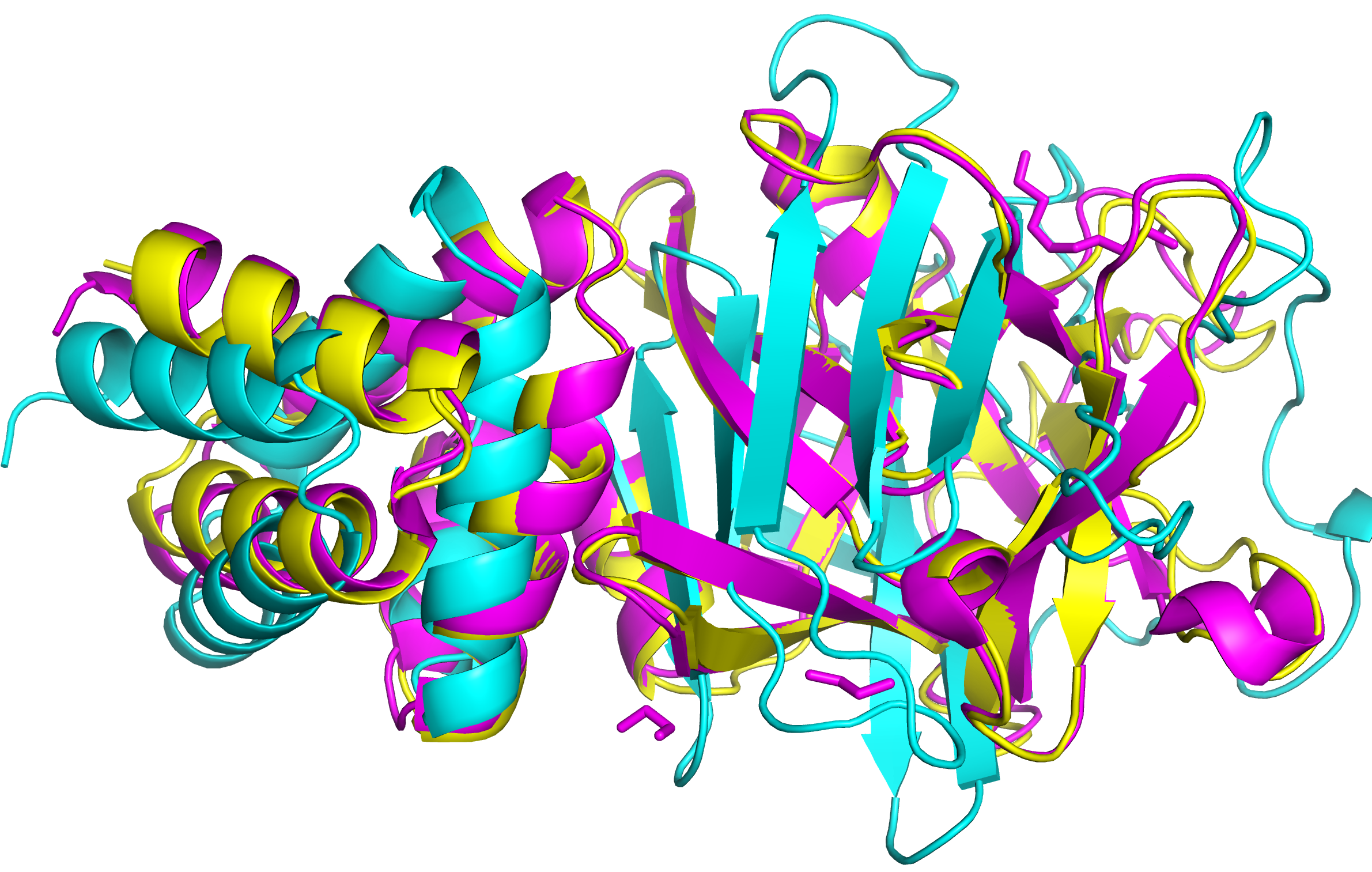} \\
    \multicolumn{2}{c}    {\includegraphics[width=0.5\linewidth]{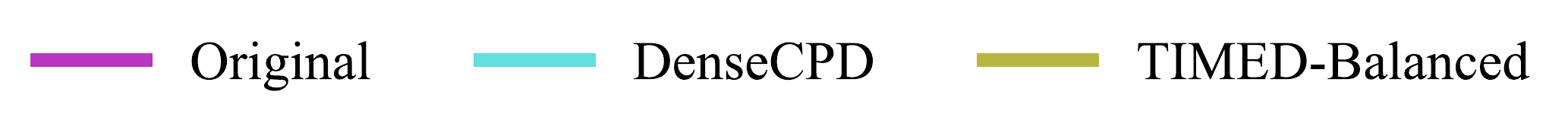}}
    \end{tabular}
    \caption{
    \label{fig:af2_densecpd_vs_timed}
    Comparison of two selected ground truth structures (pink) with those obtained using AF2 on predictions from DenseCPD (cyan) and TIMED (yellow). For 3CXB (left), RMSD is 1.34 \r{A} (DenseCPD) and  14.65\r{A} (TIMED). For 4EFP (right), it is   16.38 \r{A} (DenseCPD) and 0.65\r{A} (TIMED). }
  
\end{figure}

\section{Prediction Bias for TIMED and ProDCoNN (balanced and unbalanced)}

The composition of amino acids in proteins is not uniformly distributed. Therefore, we investigated the effect of balancing amino acid distribution in the training set. 
\begin{figure}[H]
    \centering

    \begin{tabular}{@{}c@{}c@{}}

    \includegraphics[width=0.5\linewidth]{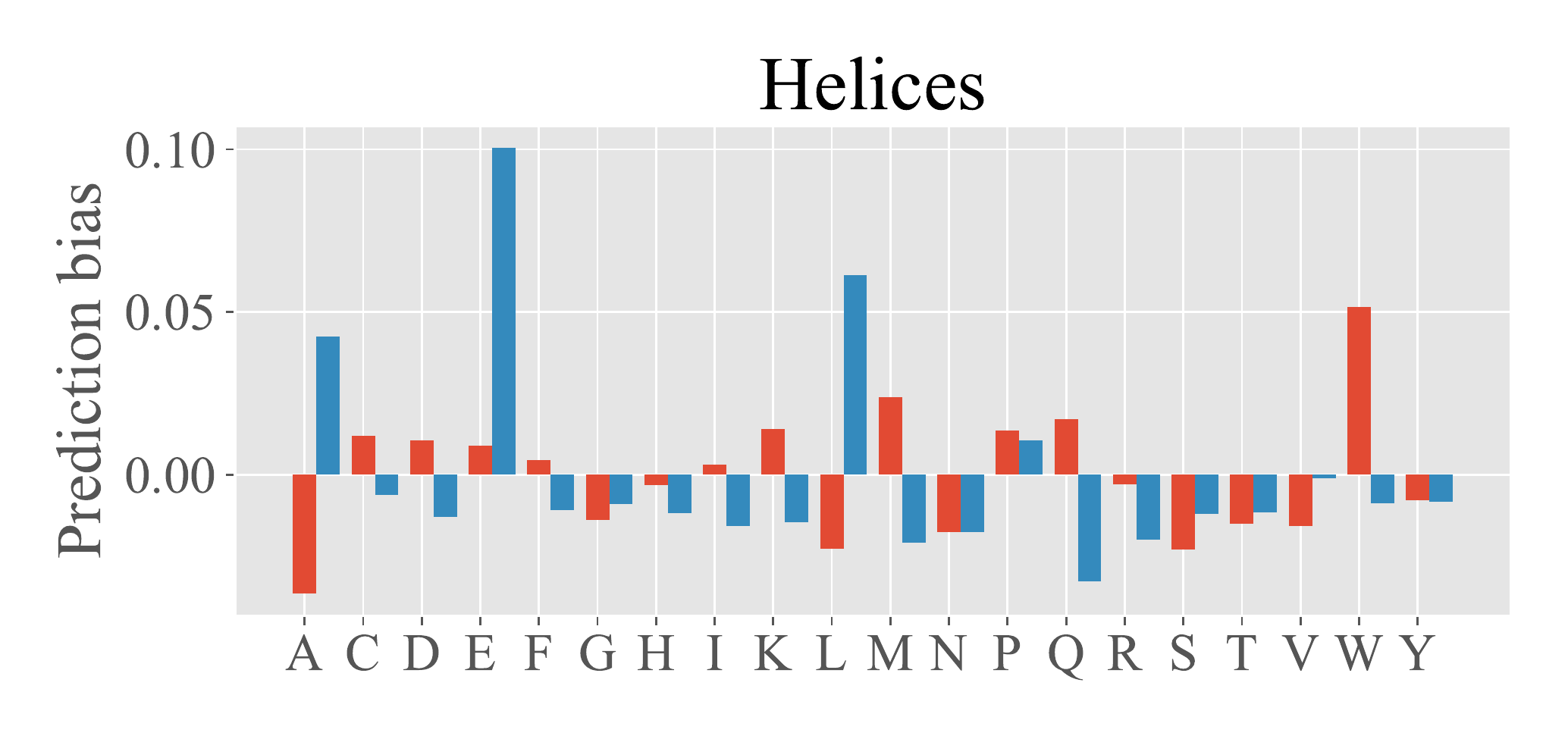} &
    \includegraphics[width=0.5\linewidth]{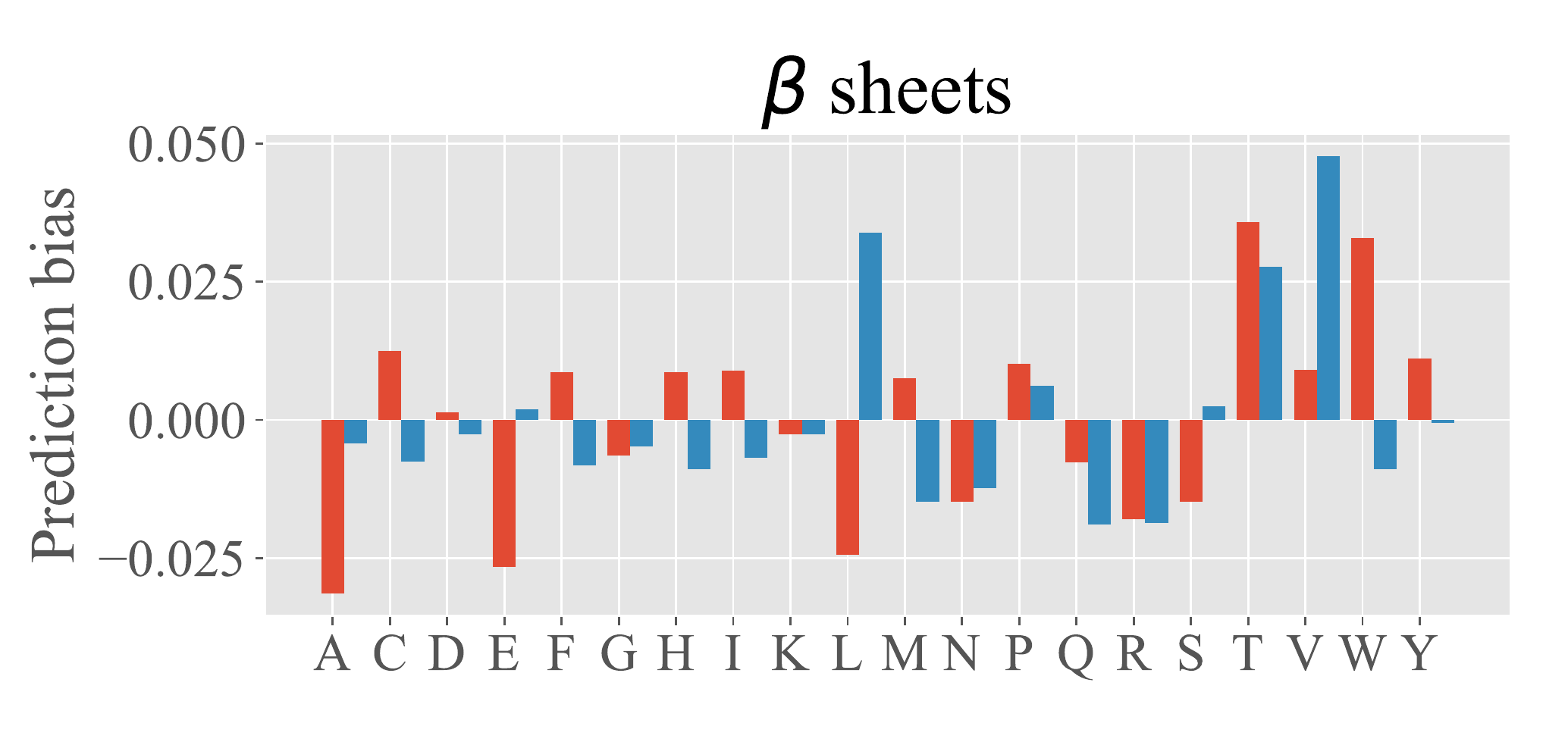}  \\
    \multicolumn{2}{c}    {\includegraphics[width=0.4\linewidth]{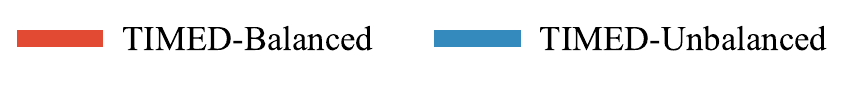}}
    \end{tabular}
    \caption{
    \label{fig:prediction_bias_alpha}
    Prediction bias comparison for TIMED-Balanced (red) and TIMED-Unbalanced models across all the structures in the benchmark and for each type of residue. The left plot represents bias on $\alpha$-helical structures, while the plot on the right is for $\beta$-sheets. Prediction bias is calculated as deviation of the predictions from the real frequency of residues in the benchmark structures.}
  
\end{figure}
\begin{figure}[H]
\centering
        \includegraphics[width=0.8\textwidth]{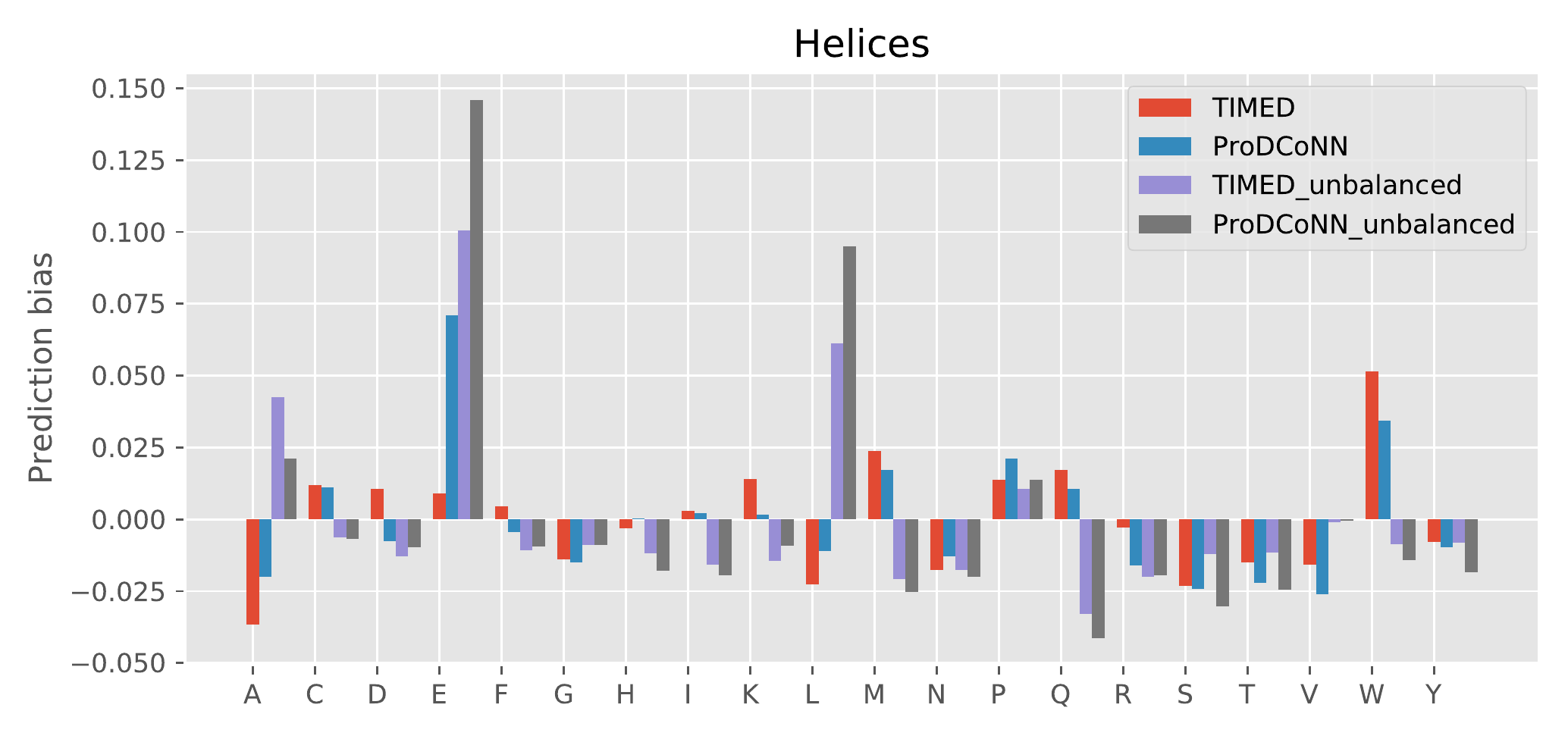}
        \includegraphics[width=0.8\textwidth]{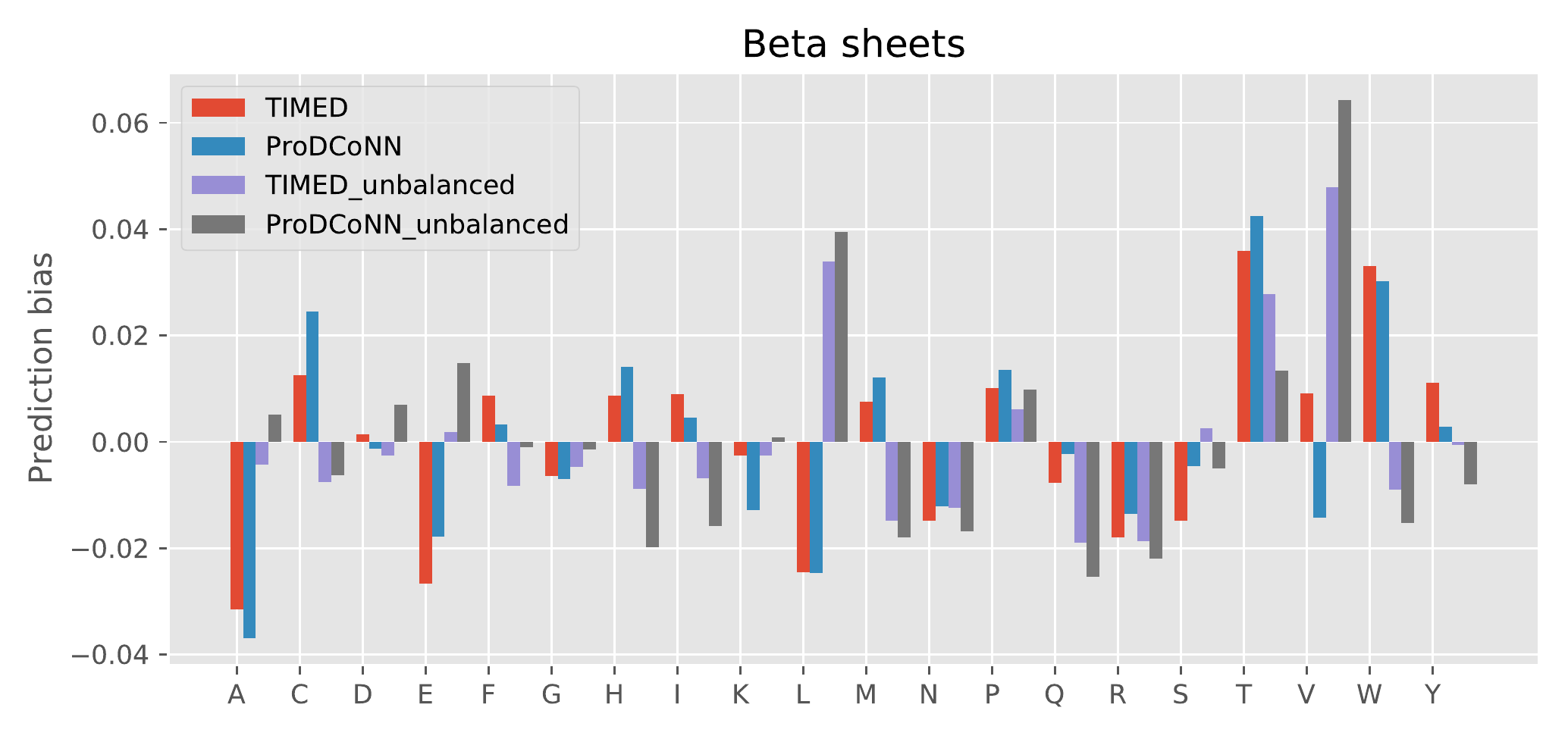}
    \caption{
    \label{fig:prediction_bias_alpha2}
    Prediction bias comparison for TIMED and ProDCoNN models both balanced and unbalanced versions.}

\end{figure}

\section{Comparison of Number of structures at each resolution for each model}

\begin{figure}[H]
 \centering
    \begin{tabular}{@{}c@{}}
         \includegraphics[width=\linewidth]{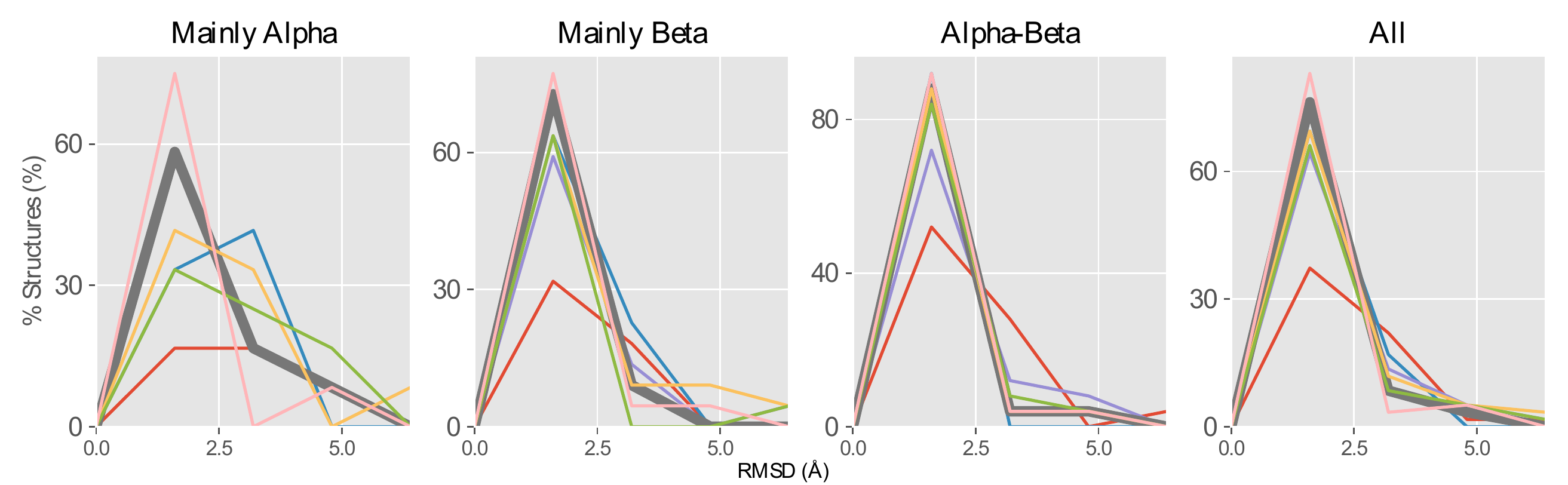}\\
     \includegraphics[width=\linewidth]{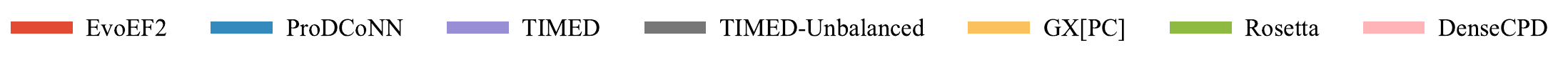}
     \end{tabular}
     \caption{Percentage of structures at each RMSD under 6.4 \r{A}, for each model and for each protein fold. Here, we used bins of 1.6 \r{A} which is the average RMSD AlphaFold2 got in CASP14.}
   \label{fig:af2_rmsd_results16}
\end{figure}

\begin{figure}[H]
 \centering
    \begin{tabular}{@{}c@{}}
         \includegraphics[width=\linewidth]{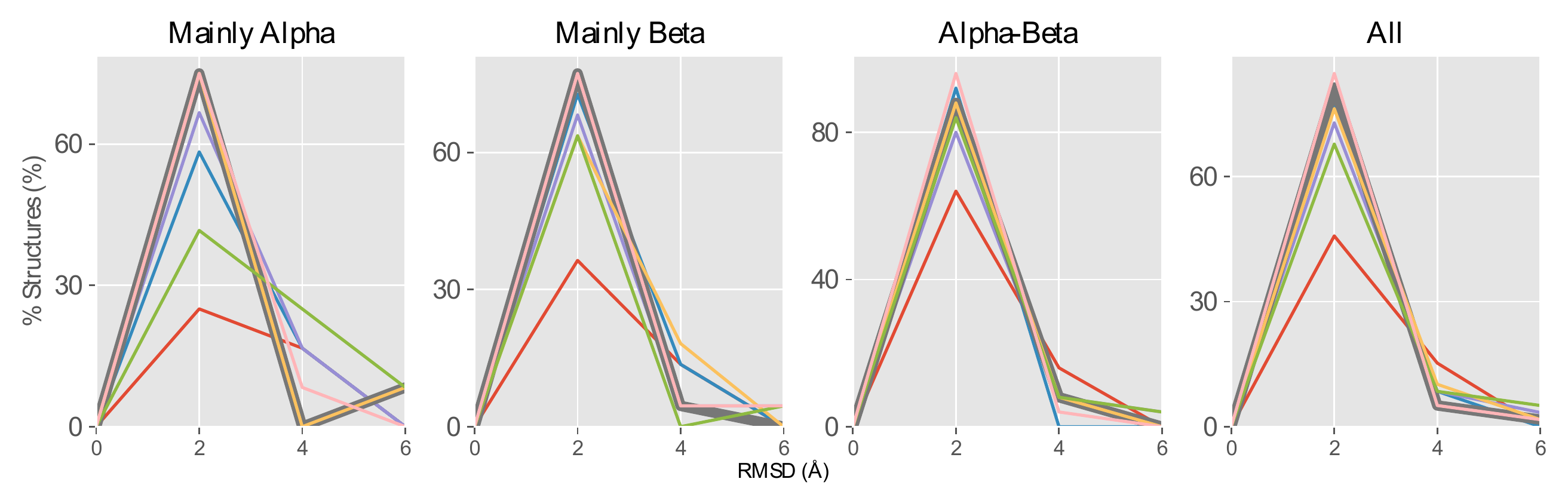}\\
     \includegraphics[width=\linewidth]{sup_img/legend}
     \end{tabular}
     \caption{Percentage of structures at each RMSD under 6 \r{A}, for each model and for each protein fold. We used bins of 2 \r{A} as this was the mean resolution across the input structures of the benchmark. }
   \label{fig:af2_rmsd_results}
\end{figure}

\begin{figure}[H]
 \centering
    \begin{tabular}{@{}c@{}}
         \includegraphics[width=\linewidth]{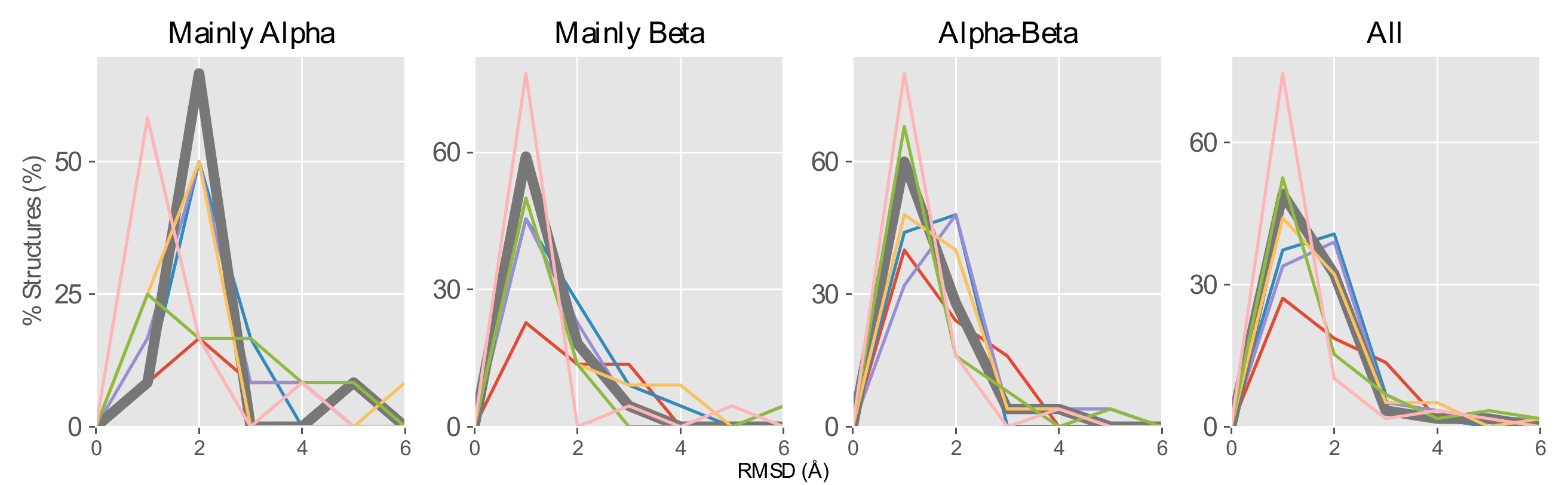}\\
     \includegraphics[width=\linewidth]{sup_img/legend}
     \end{tabular}
     \caption{Percentage of structures at each RMSD under 6 \r{A}, for each model and for each protein fold. Here, we used bins of 1 \r{A} to show the breakdown of performance below 2 \r{A}.}
   \label{fig:af2_rmsd_results2}
\end{figure}

Here, we show the \% of structure obtained at each resolution. The average RMSD for AlphaFold2 in the CASP 2020 challenge was approximately 1.6 \r{A} \cite{deepmind_2020}. We therefore show the number of structures at each bin of 1.6 \r{A} in Figure \ref{fig:af2_rmsd_results16} and  2 \r{A} in Figure \ref{fig:af2_rmsd_results}. For completeness we also show the results for 1 \r{A} in Figure \ref{fig:af2_rmsd_results2}, although these are below the margin of error and are likely not significant. 

\section{ProteinSolver Distance Matrix with and without Poly-Glycine Input}

Protein Solver Distance Matrices using 1QYS protein with and without sidechain atoms (labels). ProteinSolver claims to use the heaviest atom in the residue and as confirmed by private conversation received on June 5th at 20:59, they ``include both backbone \textbf{and side chain atoms} when calculating nearest distances''.
    
The side chains atoms determine the identify of the amino acids and are therefore labels. In a truly \textit{de novo} design setting, only an empty backbone (without side chains would be available. 

The macro-recall performance dropped from 36.6 to 13.3 when using an empty backbone (poly-glycine) input. The performance drop is only observed for ProteinSolver while it remains constant for all other models.

\begin{figure}[H]
  \includegraphics[width=1\linewidth]{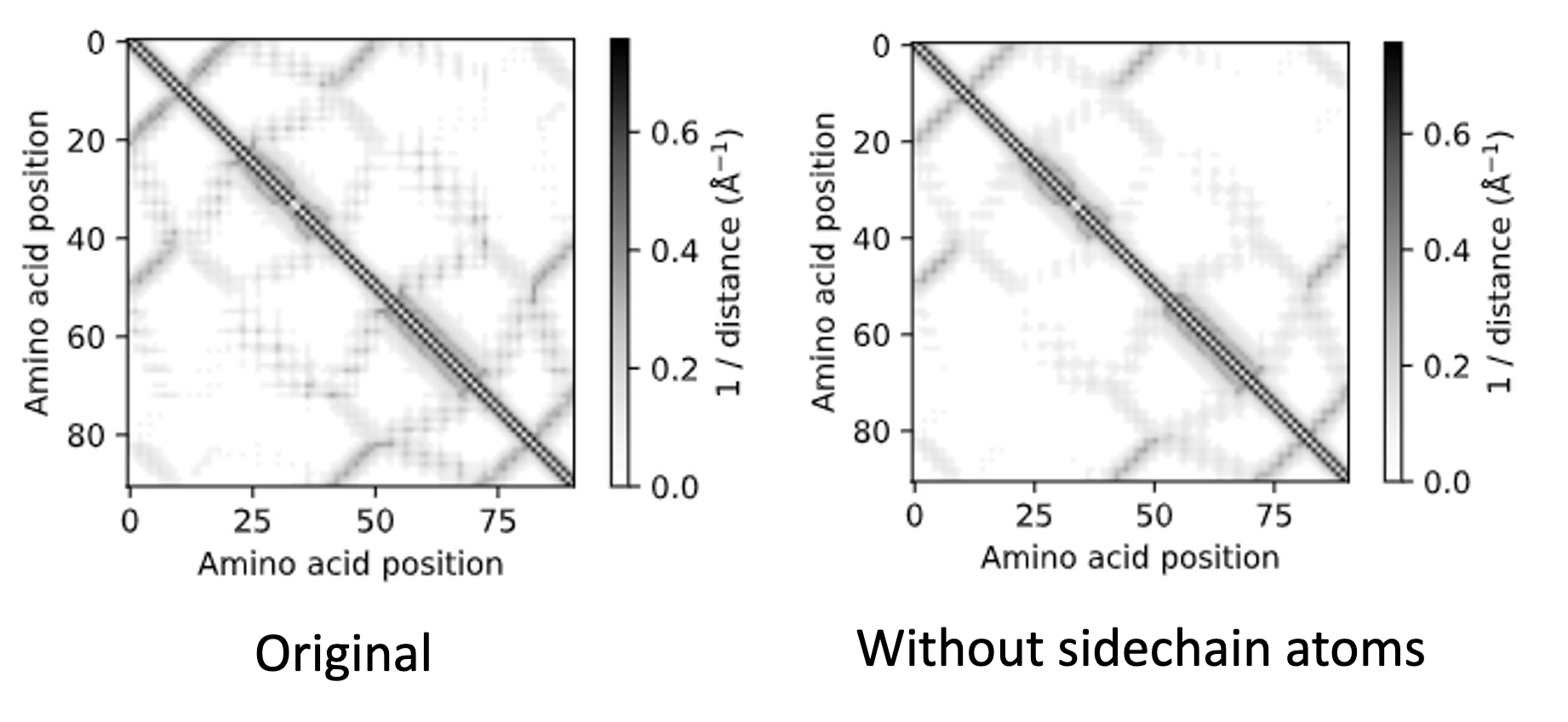}
    \caption{
    The distance matrix changes if side-chains (labels) are included in the .pdb input. The two distance matrices should be identical as only the empty backbone atoms should be used, as they are the only ones available in \textit{de novo} settings, meaning it probably presents label leakage. 
    }
  \label{fig:protein_solver}
\end{figure}

\section{Input Data Pipeline}
\begin{figure}[H]
 \centering
   \includegraphics[width=1\linewidth]{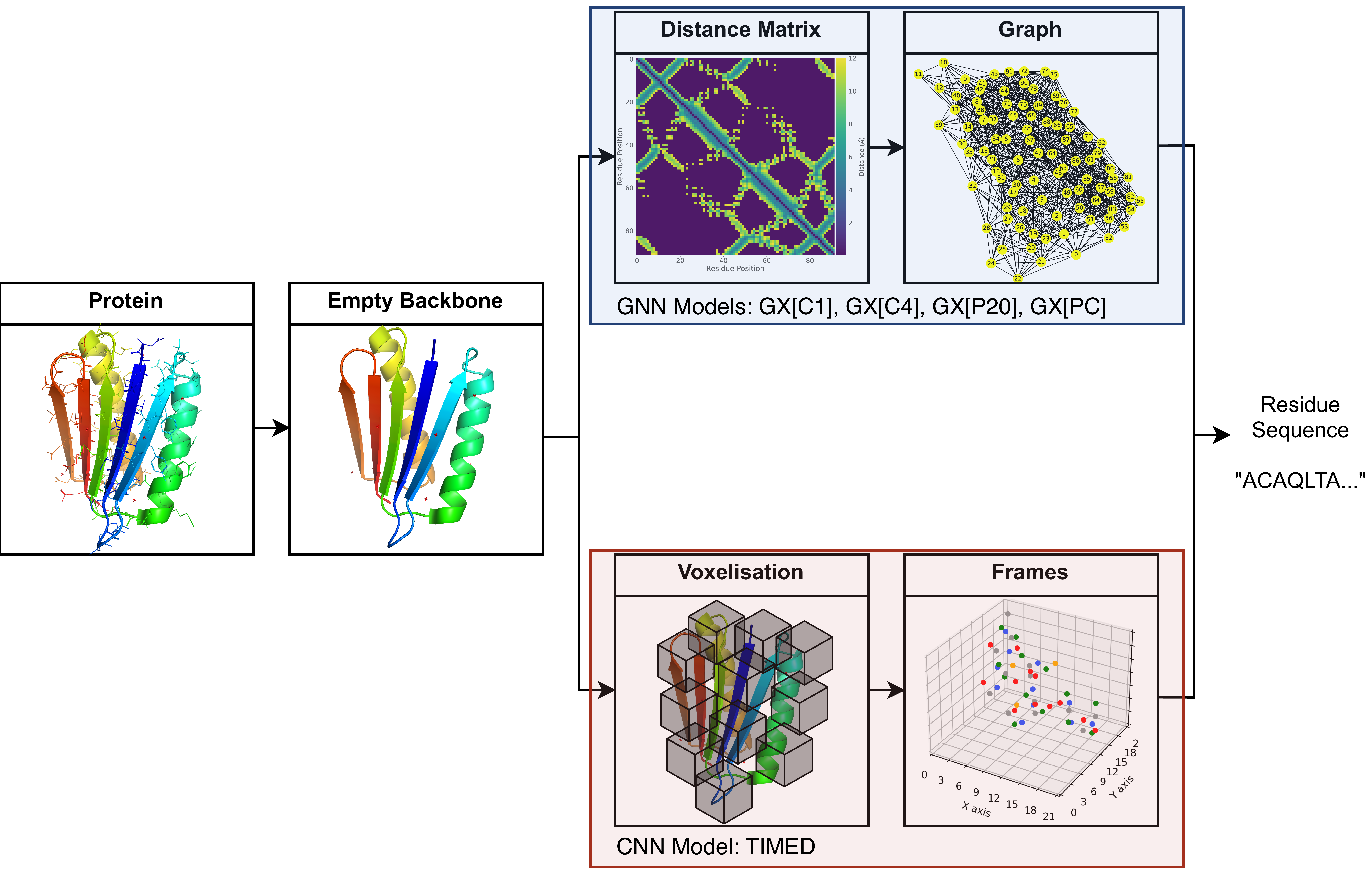}
     \caption{Illustration of the data pipeline. High-quality 3D structures of proteins are obtained from a database. The side-chains of each residues are removed so to produce an empty backbone. The GNN models calculate the distances between each residue in the protein to produce a distance matrix which is used for the production of a graph. The CNN model, on the other hand, voxelises areas of space (``frame'') around each residue, with the C$\alpha$ at the center of it. Both models predict the identity of the side-chains of the residues giving a sequence of predicted residues to obtain the input 3D structure.}
   \label{fig:data_pipeline}
 \end{figure}

\section{Physics Models Commands}\label{physics}

We evaluated two state-of-the-art physics-based methods: \textbf{EvoEF2}~\cite{Huang2019} and \textbf{Rosetta}~\cite{alford_rosetta_2017}. We used the following commands to run fixed backbone design protocols:

EvoEF2:  ./EvoEF2 --command=ProteinDesign --ppint --design\_chains=B --pdb=structure.pdb to design chain B from structure.pdb.

Rosetta (version 3.12): ./fixbb.static.linuxgccrelease -s structure.pdb -linmem\_ig 10 -ignore\_unrecognized\_res -resfile file.txt. Resfile was used to select specified chain.

\section{Benchmark structures }
Our benchmark set contains 595 protein structures spanning 40 protein architectures.

\textbf{PDB code + chain:} 1xg0C, 3g3zA, 3rf0A, 4i5jA, 2ptrA, 3f0cA, 4a5uB, 2p57A, 2q0oC, 6er6A, 1h32A, 3e3vA, 3cxbA, 1dvoA, 5dicA, 2bnmA, 4pfoA, 2ebfX, 3giaA, 1a41A, 3cexA, 4ebbA, 3jrtA, 3wfdB, 4v1gA, 3qb9A, 3abhA, 3nvoA, 2o1kA, 5x56A, 2ra1A, 4adzA, 2p6vA, 3k4iA, 4lctA, 4adyA, 4zhbA, 4p6zG, 4nq0A, 3dadA, 2vq2A, 4dloA, 2of3A, 4y5jA, 2pm7A, 2hr2A, 3ro3A, 3bqoA, 3ut4A, 2yhcA, 4k6jA, 3iisM, 5agdA, 2fbaA, 3e7jA, 1v7wA, 3a0oA, 4wu0A, 4ozwA, 4cj0A, 1gxmA, 5m7yA, 4fnvA, 5gzkA, 4ayoA, 3wkgA, 3vsnA, 2jg0A, 4j5tA, 4ktpA, 4mqwA, 5lf2A, 5mriA, 5ol4B, 1bx7A, 3ca7A, 3tvjA, 3tbdA, 1uzkA, 5bq8A, 3klkA, 1b8kA, 1v6pA, 4hquA, 4k8wA, 6a2qA, 1lpbA, 3hrzB, 6fmeB, 2aydA, 2ra8A, 4fzqA, 3d4uB, 3wwlA, 2r01A, 1lslA, 3f3fC, 2q4zA, 2de6A, 3d9xA, 2hjeA, 3mcbB, 2y8nB, 3witA, 1ya5T, 2dyiA, 3kyfA, 2v76A, 2e12A, 1g3pA, 4o06A, 3fb9A, 2p38A, 1igqA, 4hhvA, 3teeA, 5j3tA, 5h3xA, 3zbdA, 5d7uA, 5zcjC, 5u1mA, 1wthD, 4rg1A, 1kt6A, 2ja9A, 1i4uA, 4i86A, 1o7iA, 1x8qA, 2ichA, 3dzmA, 3n91A, 1luzA, 4lqzA, 4i1kA, 5xlyB, 3a35A, 3tdqA, 4mxtA, 3wjtA, 3buuA, 3ksnA, 2w7qA, 2yzyA, 4z48A, 3bk5A, 4qa8A, 2byoA, 3bmzA, 4egdA, 4joxA, 3h6jA, 2bhuA, 1pmhX, 6ggrA, 4dqaA, 4v2bA, 4weeA, 2w07B, 4r9pA, 2r2cA, 2r0hA, 4aqoA, 4luqC, 3iagC, 1k5nA, 2ygnA, 3bwzA, 4fmrA, 1njhA, 4hi6A, 1pkhA, 1gp0A, 3q1nA, 2ag4A, 2v3iA, 3ty1A, 1gprA, 3aihA, 4c4aA, 1tulA, 4a02A, 4c08A, 4maiA, 1jovA, 3wmvA, 2fdbM, 1dqgA, 1xzzA, 6i18A, 4i4oA, 4efpA, 5yh4A, 3h6qA, 5bowA, 5vi4A, 2vxtI, 3vwcA, 4lo0C, 1sr4C, 2dpfA, 3dzwA, 3a0eA, 1xd5A, 4h3oA, 4tkcA, 5j76A, 3mezC, 4gc1A, 1b2pA, 4le7A, 4oitA, 6b0gE, 1z1yB, 1vmoA, 2gudA, 4r6rE, 5krpC, 5v6fA, 4pitA, 6flwA, 4ddnD, 3apaA, 5gvyA, 1c3mA, 4mq0A, 3wocA, 3aqgA, 3towA, 2qp2A, 1nykA, 2bmoA, 2gbwA, 1rfsA, 4aivA, 3gkeA, 2nwfA, 1jm1A, 2qpzA, 5cxmB, 3dqyA, 3d89A, 2b1xA, 4qdcA, 2q3wA, 3c7xA, 1genA, 1itvA, 3s18A, 4rt6B, 3cu9A, 3wasA, 6ms3B, 6frwA, 3k1uA, 5aycA, 5c0pA, 4n1iA, 3r4zA, 1tl2A, 4u6dA, 1oygA, 4qqsA, 3qz4A, 5a8cA, 4pvaA, 3kstA, 5flwA, 6gy5A, 1cruA, 1suuA, 3o4pA, 2p4oA, 4mzaA, 5gtqA, 3dr2A, 3dasA, 3g4eA, 2fp8A, 5hx0B, 1npeA, 1s1dA, 2zwaA, 3a72A, 2zb6A, 3scyA, 3b7fA, 3al9A, 3o4hA, 4pxwA, 4wk0A, 2w18A, 5em2A, 1sq9A, 1xipA, 4h5iA, 1jofA, 5ic7A, 5k19A, 6e1zA, 2z2nA, 6e4lA, 6fkwA, 6damA, 1flgA, 4cvbA, 4mh1A, 1z68A, 2z3zA, 4q1vA, 1xfdA, 5d7wA, 1kapP, 3laaA, 1p9hA, 3ultA, 3s6lA, 2xqhA, 4dt5A, 5m5zA, 5lw3A, 1k5cA, 1k4zA, 2ntpA, 3bh7B, 2j8kA, 2vfoA, 3n6zA, 1hf2A, 2x3hB, 1rmgA, 6mfkA, 1l0sA, 2xt2A, 5nzgA, 3kweA, 2w7zA, 1lktA, 3facA, 3pyiB, 2casA, 1gppA, 3maoA, 1ut7A, 1hxrA, 1t61A, 4qjvA, 3lywA, 3dalA, 5hqhA, 3u7zA, 3r90A, 1tp6A, 3s9xA, 2ex5A, 3gbyA, 5kvbA, 2cu3A, 1c1yB, 5f6rA, 4a6qA, 2w56A, 4lqbA, 4oobA, 3oajA, 3n8bA, 3jumA, 2prxA, 5b1rA, 1ewfA, 4m4dA, 2obdA, 6baqA, 1usuB, 3e8tA, 3aotA, 2rckA, 3l6iA, 3uv1A, 3bqwA, 5mprA, 1kkoA, 4cd8A, 1vd6A, 2g0wA, 4lanA, 3s83A, 2v3gA, 3fkrA, 4z0gA, 3sggA, 5zjbA, 2xfrA, 4g8tA, 5n6fA, 1muwA, 2qhqA, 3h35A, 3kluA, 3fn2A, 2od6A, 1kcfA, 3nlcA, 2zw2A, 4ftxA, 3u2aA, 2hiqA, 1xkpC, 6ih0A, 5c12A, 1w4rA, 3c0fB, 3nbmA, 2r6zA, 5hxdA, 1chdA, 3do8A, 3gohA, 1n0eA, 2q82A, 5kxhA, 3oqiA, 2x4lA, 3d3kA, 3l46A, 2fkcA, 5jphA, 3nytA, 3rhtA, 3dkrA, 2psbA, 1tc5A, 3vrdB, 2je3A, 3g5sA, 1jkeA, 4at0A, 1vi4A, 4u8pC, 4ntcA, 5ipyA, 5nakA, 4z24B, 4opcA, 5cdkA, 2b0aA, 4n2pA, 1j5uA, 1vzyB, 1vq0A, 4ipuA, 4dq9A, 4jtmA, 3gs9A, 3adyA, 3mi0A, 3ib7A, 3g91A, 1vr7A, 4zx2A, 1ds1A, 3zwfA, 1hq0A, 3hbcA, 3p8kA, 1wraA, 3t91A, 3c9fA, 2imhA, 1um0A, 5y0mA, 5u4hA, 3zh4A, 3swgA, 5ujsA, 3nvsA, 2o0bA, 2pqcA, 3slhA, 1rf6A, 4n3pA, 5bufA, 3rmtA, 4fqdA, 1ud9A, 1t6lA, 1rwzA, 3ifvA, 1iz5A, 3lx2A, 1u7bA, 5tupA, 5h0tA, 5v7mA, 3fdsC, 3aizA, 1b77A, 3p91A, 1dmlA, 3hslX, 2z0lA, 6nibA, 2jerA, 1xknA, 1zbrA, 3hvmA, 1jdwA, 5wpiA, 1g61A, 1h70A, 5m3qA, 1ynfA, 3wn4A, 1io0A, 4rcaB, 4fcgA, 4ecoA, 3wpcA, 4im6A, 4cnmA, 5hzlB, 4fs7A, 2xwtC, 3e4gA, 4wp6A, 5il7A, 1z7xW, 4u7lA, 6fg8A, 2wfhA, 2fy7A, 5wwdA, 1j3aA, 1omzA, 3emfA, 1xw3A, 3h4rA, 3essA, 1o22A, 4ktbA, 1jh6A, 3n08A, 5tsqA, 3e9vA, 4j7hA, 1i4jA, 2wnfA, 3v1aA, 3coqA, 2f60K, 4zgmA, 1i7wB, 6g6kA, 1pbyC, 1a92A, 3alrA, 2wjvD, 2a26A, 1devB, 4l0nA, 4ayaA, 3zxcA, 4pkfB, 2b1yA, 4dncD, 4jpnA, 4e18B, 3vepX, 3v4yB, 1xawA, 1ykhA, 2p64A, 6bscB, 2z3xA, 4uzzB, 3thfA, 1wq6A, 4ke2A, 4lhfA, 2v66B, 3lczA, 2h4oA, 4wjwA, 3kvpA, 3e56A, 3bk3C, 2ds5A, 3zoqB, 3nfgB, 4ksnA, 3ua0A, 3nrtA, 4a9aC, 6hikL

\section{TIMED CNN Architecture}

\begin{figure}[H]
 \centering
  \includegraphics[max size={\textwidth}{0.9\textheight}]{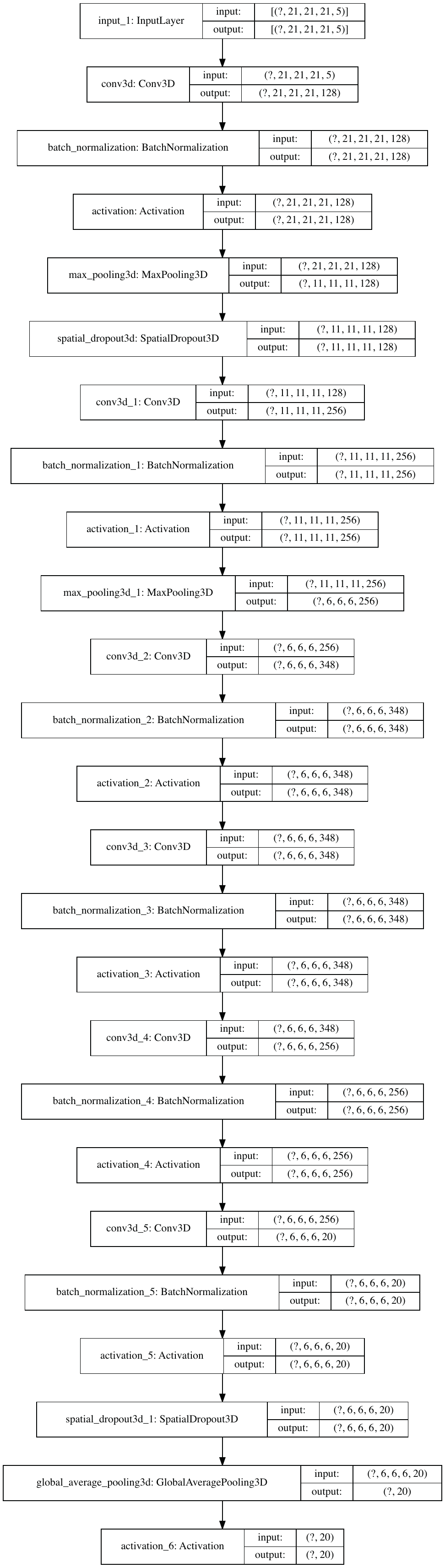}
    \caption{Architecture of the TIMED Convolutional Model.}
  \label{fig:cnn_architecture_fig}
\end{figure}

\section{GX[PC] GNN Architecture}

Similarly to ProteinSolver we create a graph structure for each protein. We calculate the distance between the C$\alpha$ of all residues in the empty backbone to produce a distance matrix for each protein. The distance matrices were then loaded into SciPy's sparse matrix \cite{Virtanen2020} and Deep Graph Library (DGL) \cite{wang2019dgl} to produce a graph structure. Each residue in the protein sequence is represented by a node. An edge connects two nodes if the distance between the two is below the distance threshold, for example 12 \r{A}, which was obtained from ProteinSolver \cite{strokach_fast_2020}. We also tried different ranges of distances ranging from 4 to 10 \r{A}.

The nodes in the graphs were given different node features:

- \textbf{C4}: Four atomic coordinates of the node, C, N, O, and C$\alpha$. 

- \textbf{P20}: The output 20-dimensional probability of TIMED (CNN), for that specific node.

- \textbf{PC}: Both C4 and P20 in a 24-dimensional array.

We used a 70-30 train-validation split, again  excluding all benchmarking structure. As with Aposteriori, we release the code open-source together with our models.

To circumvent DGL's limitation of not being able to save datasets larger than RAM, we modified the \textit{DGLDataset} class to save (and load) batches of graphs into multiple files as well as supporting multiprocessing. Data Loading in batches significantly affects the training time.

The best GX model was trained with a distance threshold of 12 and 10 layers with the following architecture:

10 layers with SAGEConv using the mean aggregation function. Each layer is followed by activation with LeakyRelu and Dropout rate of 0.4, except for the final layer.

\section{DenseCPD Architecture (claimed)}

\begin{figure}[H]
 \centering
  \includegraphics[max size={\textwidth}{0.9\textheight}]{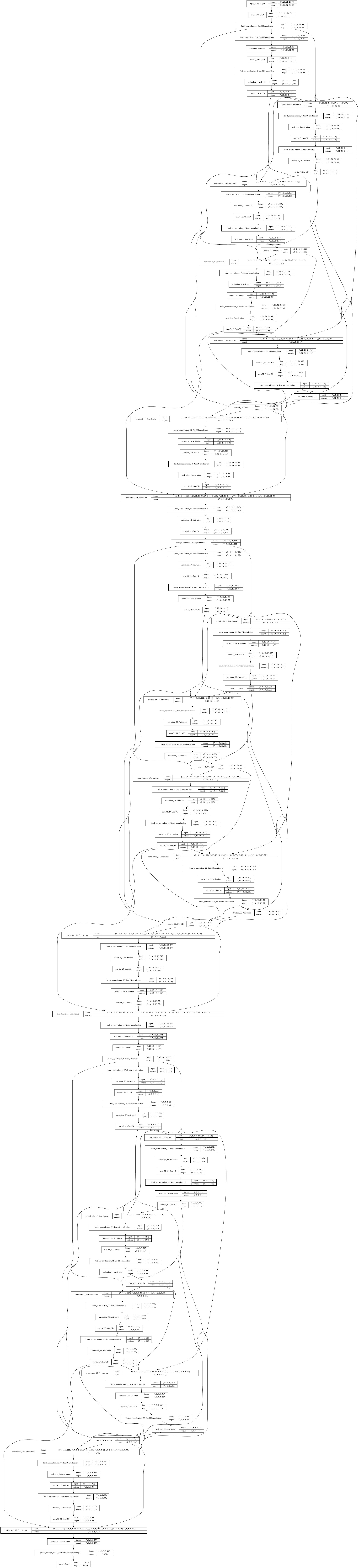}
    \caption{Replication of architecture of the DenseCPD from the paper figure.}
  \label{fig:dense_claimed_architecture_fig}
\end{figure}

\section{DenseCPD Architecture (actual)}

\begin{figure}[H]
 \centering
c  \includegraphics[max size={\textwidth}{0.9\textheight}]{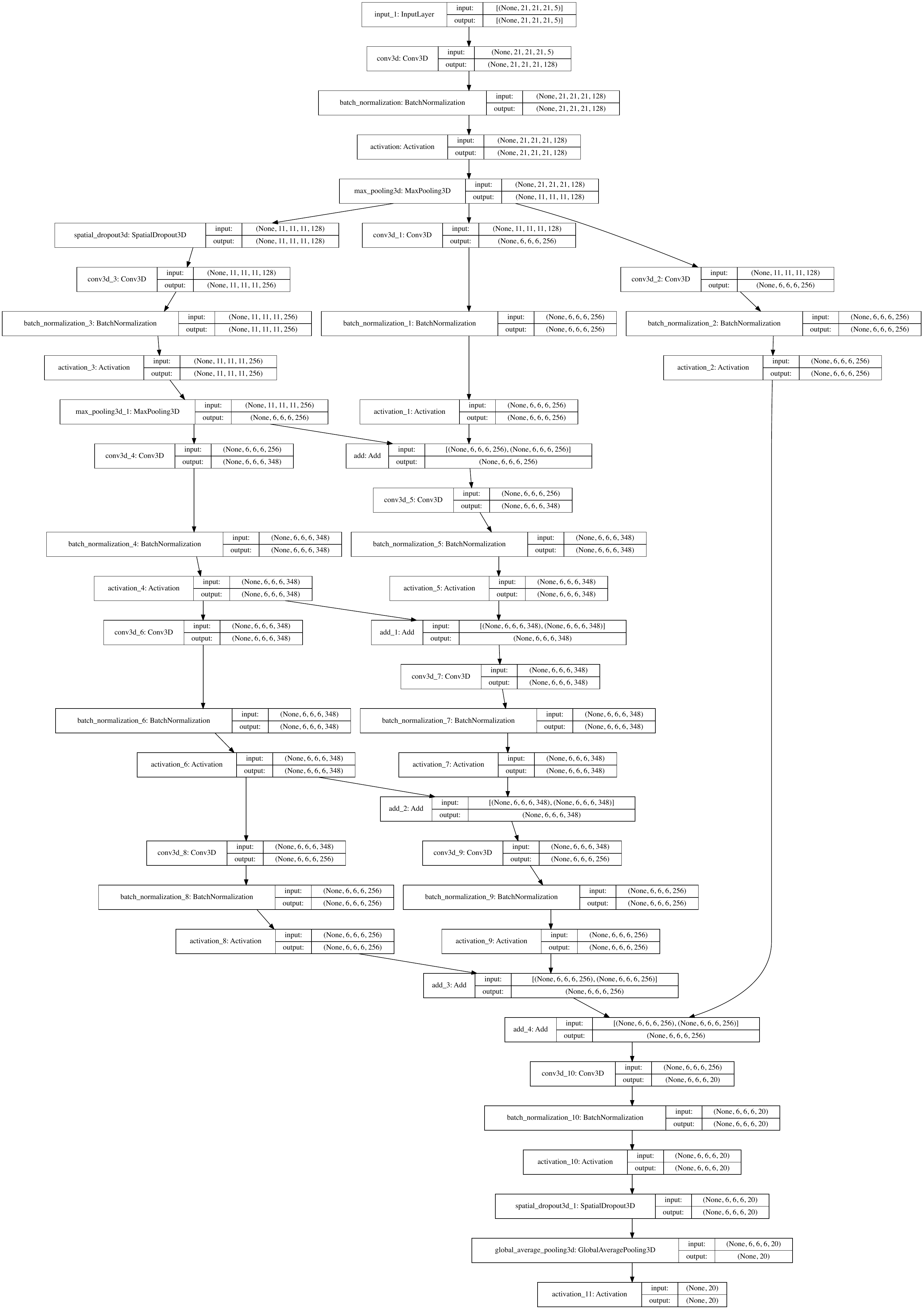}
    \caption{Architecture of the DenseCPD obtained from the model.json file from the authors.}
  \label{fig:dense_resal_architecture_fig}
\end{figure}

\bibliography{references}